\documentclass[journal,compsoc]{IEEEtran}
\usepackage{cite}
\usepackage{amsmath,amssymb,amsfonts}
\usepackage{ragged2e}
\usepackage{algorithmic}
\usepackage{graphicx}
\usepackage{textcomp}
\usepackage{xcolor}
\usepackage{hyperref}
\usepackage{cleveref}
\usepackage{subfiles}
\usepackage{soul}
\usepackage{lscape}
\usepackage{tabularray}
\UseTblrLibrary{diagbox}
\usepackage{supertabular}
\usepackage{longtable}
\usepackage{multirow}

\usepackage{pifont}
\usepackage[table,xcdraw]{xcolor}
\usepackage{tikz}
\usepackage{mathtools}
\usetikzlibrary{arrows.meta}
\usepackage[normalem]{ulem}
\useunder{\uline}{\ul}{}
\usepackage{booktabs} 
\usepackage{mdframed}
\usepackage{physics}

\newcommand{\new}[1]{{\color{black}{#1}}}

\def\BibTeX{{\rm B\kern-.05em{\sc i\kern-.025em b}\kern-.08em
T\kern-.1667em\lower.7ex\hbox{E}\kern-.125emX}}
\crefformat{section}{\S#2#1#3} 
\crefformat{subsection}{\S#2#1#3}
\crefformat{subsubsection}{\S#2#1#3}

\begin{document}

\title{When Federated Learning Meets Quantum Computing: Survey and Research Opportunities}

\author{Aakar Mathur, Ashish Gupta,~\IEEEmembership{Member,~IEEE}, and~Sajal K. Das,~\IEEEmembership{Fellow,~IEEE} 
\IEEEcompsocitemizethanks{\IEEEcompsocthanksitem{Aakar Mathur, Ashish Gupta are with the Department of Computer Science and Engineering, BITS Pilani Dubai Campus, Dubai, UAE.\\ Sajal K. Das is with the Department of Computer Science, Missouri University of Science and Technology, Rolla, USA.} \protect\\
E-mail: \{f20220120, ashish\}@dubai.bits-pilani.ac.in, sdas@mst.edu}
\thanks{}}

\IEEEtitleabstractindextext{
\justifying
\begin{abstract}
Quantum Federated Learning (QFL) is an emerging field that harnesses advances in Quantum Computing (QC) to improve the scalability and efficiency of decentralized Federated Learning (FL) models. This paper provides a systematic and comprehensive survey of the emerging problems and solutions when FL meets QC, from research protocol to a novel taxonomy, particularly focusing on both quantum and federated limitations, such as their architectures, Noisy Intermediate Scale Quantum (NISQ) devices, and privacy preservation, so on. With the introduction of two novel metrics, qubit utilization efficiency and quantum model training strategy, we present a thorough analysis of the current status of the QFL research. This work explores key developments and integration strategies, along with the impact of QC on FL, keeping a sharp focus on hybrid quantum-classical approaches. The paper offers an in-depth understanding of how the strengths of QC, such as gradient hiding, state entanglement, quantum key distribution, quantum security, and quantum-enhanced differential privacy, have been integrated into FL to ensure the privacy of participants in an enhanced, fast, and secure framework. 
Finally, this study proposes potential future directions to address the identified research gaps and challenges, aiming to inspire faster and more secure QFL models for practical use.
\end{abstract}
}

\maketitle
\section{Introduction}
\label{sec:introduction}

\IEEEPARstart{F}{ederated} Learning (FL) represents a decentralized paradigm in Machine Learning (ML), introduced by Google researchers~\cite{mcmahan2017communication}. Unlike traditional centralized approaches, FL facilitates the collaborative training of a global model by multiple clients, preventing the need for local data to be transferred to a central server~\cite{FL1}. Clients undertake local training of models, sharing solely the model parameters with the central server for aggregation~\cite{FL2}. This approach ensures data privacy while facilitating learning from diverse datasets across various devices. 
Moreover, it capitalizes on data heterogeneity, enabling model training even in instances where data among clients are non-Independent and Identically Distributed (non-IID). Nevertheless, FL confronts technical challenges, including heterogeneity, asynchronous updates, and communication efficiency between devices and servers. As FL progresses, it is becoming an indispensable framework in privacy-preserving applications across a multitude of industries ~\cite{FLApp2,FLApp1,FLApp3}.

\begin{figure}
    \centering
    \includegraphics[width=1\linewidth]{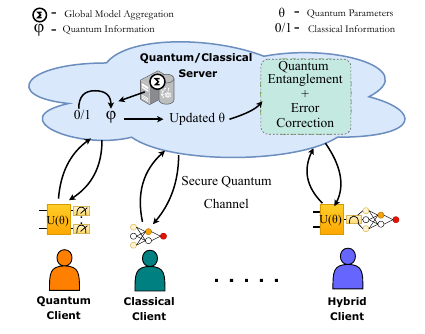}
    \caption{A QFL framework with different types of clients.}
    \label{fig:qfl-framework}
\end{figure}

From a computing aspect, Quantum computing (QC) is a rapidly advancing domain that employs the principles of quantum mechanics to execute computations beyond the efficient capabilities of classical computers~\cite{QC1}. Central to QC is the notion of quantum bits,  commonly known as {\em qubits}, which, in contrast to classical bits, possess the ability to exist in superposition(ability of a system to exist in multiple states simultaneously). This characteristic, in conjunction with entanglement~\cite{901921,9779492} and quantum  interference~\cite{philippidis1979quantum,9716782}, facilitates quantum computers in executing parallel computations at an unprecedented scale, rendering them particularly adept at addressing intricate optimization problems, cryptographic challenges, and simulations of quantum physical systems~\cite{QC3,QC4,QC5,QC6}. QC holds the promise to transform disciplines such as cryptography, materials science, and particularly ML. Quantum algorithms such as Shor's algorithm~\cite{QC7} for factorization of large numbers and Grover's search algorithm~\cite{QC8}, illustrate the potential of QC to surpass classical systems in specific applications. 

Quantum Federated Learning (QFL) combines FL and QC to meet the growing computational demands of distributed systems and enhance privacy~\cite{QFL1}. Traditional FL struggles with scalability and privacy, especially with large data sets and complex models. QC can address these by speeding up tasks like model optimization and parameter aggregation~\cite{QFL2, QFL3}. QFL framework (as illustrated in Fig.~\ref{fig:qfl-framework}) is promising for real-time data processing in edge computing and healthcare~\cite{QFL4}, offering robust, fast, and secure ML models by merging FL with efficient quantum algorithms such as Quantum Approximate Optimization Algorithms (QAOA) ~\cite{QFL5} on Noisy Intermediate Scale Quantum (NISQ) devices \cite{Preskill2018}.

\noindent $\bullet$ {\bf Motivation:} While several surveys on QFL exist, considerable research gaps persist, necessitating a more rigorous examination. The early work in~\cite{narottama2020quantum} introduced a two-tier QFL model for wireless communications, but it lacks empirical validation and overlooks practical concerns such as quantum noise, scalability, and communication overhead. Later studies, such as~\cite{huang2022quantum, ren2023towards, gurung2023quantum}, attempted to categorize QFL techniques but could not sufficiently establish the foundational connections between classical FL and quantum frameworks. Research efforts in~\cite{chehimi2024foundations, quy2024from, QFL-type, javeed2023quantum} have made strides in outlining the theoretical foundations of QFL, yet they do not incorporate evaluation methodologies for quantum communication constraints and real-world applicability. A recent study~\cite{quy2024from} explored QFL in the context of space-air-ground integrated networks but failed to address key hardware limitations in NISQ devices. Furthermore,~\cite{saha2024multifaceted} presented the research opportunities related to privacy aspects of QFL.
Surprisingly, none of the above studies analyzes the practical implications of QFL frameworks on NISQ devices and also fails to provide a meaningful way to categorize QFL research.  Thus, a more comprehensive and systematic study is needed to gain crucial insight into the potential types of QFL research along with their challenges. 

\noindent $\bullet$ {\bf Major Contributions:} To the best of our knowledge, this work is the first to critically analyze the integration of FL with QC. We make the following major contributions.
\begin{itemize}
    \item With a systematic research protocol, we propose a novel taxonomy of the QFL literature to gain key insights and learn about challenges.
    \item This work introduces two novel metrics (one quantitative: qubit utilization efficiency and another qualitative: quantum model training strategy) to critically examine QFL studies that include models with architectural integration, FL on NISQ devices, and privacy-preserving QFL. 
    \item The research recommendations of this study offer a strong foundation for future exploration of the dimensional intersection where FL meets QC. 
\end{itemize}
The remainder of the paper is organized as follows. Section~\ref{sec:Research Protocol} provides a systematic protocol for selecting relevant papers. Section~\ref{sec:Foundation and Taxanomy of QFL} offers a foundation for QFL and Section~\ref{sec:Taxonomy} introduces two novel evaluation metrics along with a proposal of the novel taxonomy. Sections~\ref{sec:Models with Architectural Integration},~\ref{sec:FL on NISQ Devices},~\ref{sec:Privacy-Preserving QFL}, and~\ref{sec:Miscellaneous} discuss works that fall within different categories of taxonomy. We also provide insights into experimental evaluation in Section~\ref{sec:Experimental}. Finally, Section~\ref{sec:Discussion and Future Directions} concludes the paper with recommendations for further research in QFL.

\section{Research Protocol}
\label{sec:Research Protocol}
To conduct a systematic review, we follow a research protocol, inspired by the work~\cite{gupta2020approaches}, to outline the process for searching, filtering, and selecting the relevant QFL papers. This protocol offers a structured approach to uncover notable contributions that need to be explored to gain an in-depth comprehensive understanding of the field.

\begin{table}[h]
\centering
\caption{Search query and terms used for filtering papers}
\label{tab:paper_filtering}
\resizebox{.5\textwidth}{!}{
\begin{tblr}{
  cell{1}{3} = {c},
  cell{2}{3} = {c},
  cell{2}{4} = {c},
  cell{3}{3} = {c},
  cell{3}{4} = {c},
  cell{4}{3} = {c},
  cell{4}{4} = {c},
  cell{5}{1} = {c=2}{},
  cell{5}{4} = {c},
  hlines,
  vline{-} = {1-4}{},
  vline{1,5} = {5}{},
}
Query   & Query Terms                                                                                                                                                   & {Fetched \\ papers}             & Final Papers         \\
Query 1 & {"Federated Learning" AND \\"Quantum Computing"}
& {IX = 22\\AR = 28\\GS = 53}   & {103\\(13 duplicates)} \\

Query 2 & {("Federated Learning" OR \\"Decentralized ML") AND \\("Quantum Computing" OR \\"Quantum Mechanics" OR \\"Quantum Physics")~}                                 
& {IX = 17\\AR = 34\\GS = 62~~} & {113\\(36 duplicates)} \\

Query 3 & {(("Federated Learning" OR \\"Decentralized ML") AND\\("Quantum Computing" OR\\"Quantum Mechanics" OR\\"Quantum Physics")) OR \\"Quantum Federated Learning"} 
& {IX = 35\\AR = 21\\GS =~ 49~} & {105\\(23 duplicates)} \\
        &                                                                                                                                                               & Total~                        & 321                  
\end{tblr}
}
\end{table}

\subsection{Search Strategy}
Initially, we find a set of search terms to retrieve the most relevant QFL papers from well-established academic databases, including IEEE Xplore (IX), arXiv (AR), and Google Scholar (GS), as they comprehensively cover research fields pertinent to QC, FL, and QFL. Table \ref{tab:paper_filtering} presents the query terms and the number of papers retrieved from each database. To ensure extensive coverage, we designed queries with varying levels of granularity while consistently including terms like \textit{quantum}, \textit{federated learning}, and \textit{privacy}. Specific quantum terms such as \textit{"QAOA"} and \textit{"NISQ"} were also included. The search is restricted to papers published within the past 9 years (2016 to 2025).

\subsection{Inclusion Criteria}

For a paper to be included in this review, it must satisfy the following criteria:
1) The paper must be in English.
2) The paper should be a peer-reviewed journal article, a conference proceeding, or a book chapter.
3) The studies may range from fully quantum to fully classical models using quantum algorithms.
4) Research must include practical or theoretical applications of QFL methodologies, such as quantum-enhanced differential privacy, quantum communication protocols, or QFL architectures.

\begin{figure}[h]
    \centering    \includegraphics[width=1\linewidth]{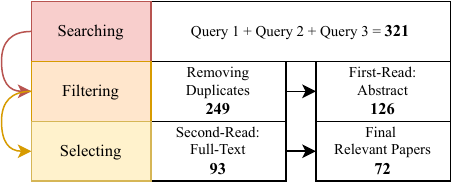}
    \caption{Details of the research protocol. 
    }
    \label{fig:selection-process}
\end{figure}

\subsection{Selection Process}

After retrieving papers using the specified query terms, duplicate papers were removed. We then screened the titles and abstracts of the remaining papers. Articles that did not meet the inclusion criteria were excluded. Subsequently, full-text screening was conducted on the remaining papers to ensure relevance and additional relevant papers were sourced from references within the selected articles. Fig. \ref{fig:selection-process} depicts the selection process, illustrating the filtering stages and the number of papers retained at each stage.

\section{Foundation of QFL}
\label{sec:Foundation and Taxanomy of QFL}
This section provides the major fundamentals of FL and QFL required to analyze the current status of the research.

\subsection{Quantum Computing}
\begin{figure}[h]
\begin{tikzpicture}[line cap=round, line join=round, >=Triangle]
  \clip(-2.19,-2.49) rectangle (7,2.58);
  \draw [shift={(0,0)}, red, fill, fill opacity=0.1] (0,0) -- (56.7:0.4) arc (56.7:90.:0.4) -- cycle;
  \draw [shift={(0,0)}, orange, fill, fill opacity=0.1] (0,0) -- (-135.7:0.4) arc (-135.7:-33.2:0.4) -- cycle;
  \draw(0,0) circle (2cm);
  \draw [rotate around={0.:(0.,0.)}, dash pattern=on 3pt off 3pt] (0,0) ellipse (2cm and 0.9cm);
  \draw (0,0)-- (0.70,1.07);
  \draw [->] (0,0) -- (0,2);
  \draw [->] (0,0) -- (-0.81,-0.79);
  \draw [->] (0,0) -- (2,0);
  \draw [dotted] (0.7,1)-- (0.7,-0.46);
  \draw [dotted] (0,0)-- (0.7,-0.46);
  \draw (-0.08,-0.3) node[anchor=north west] {$\varphi$};
  \draw (0.01,0.9) node[anchor=north west] {$\theta$};
  \draw (-1.01,-0.72) node[anchor=north west] {$\mathbf {\hat{x}}$};
  \draw (2.07,0.3) node[anchor=north west] {$\mathbf {\hat{y}}$};
  \draw (-0.5,2.6) node[anchor=north west] {$\mathbf {\hat{z}=|0\rangle}$};
  \draw (-0.4,-2) node[anchor=north west] {$-\mathbf {\hat{z}=|1\rangle}$};
  \draw (0.4,1.65) node[anchor=north west] {$|\psi\rangle$};
  \scriptsize
  \draw [fill] (0,0) circle (1.5pt);
  \draw [fill] (0.7,1.1) circle (0.5pt);
  
  \node[align=left, anchor=west] at (3,0.9) {$\mathbf{\hat{x}}$ -- x-axis reference};
  \node[align=left, anchor=west] at (3,0.6) {$\mathbf{\hat{y}}$ -- y-axis reference};
  \node[align=left, anchor=west] at (3,0.3) {$\mathbf{\hat{z}}=|0\rangle$ -- North pole (logical 0)};
  \node[align=left, anchor=west] at (2.8,0.0) {$-\mathbf{\hat{z}}=|1\rangle$ -- South pole (logical 1)};
  \node[align=left, anchor=west] at (3,-0.3) {$\theta$ -- polar angle};
  \node[align=left, anchor=west] at (3,-0.6) {$\varphi$ -- azimuthal angle};
  \node[align=left, anchor=west] at (3,-0.9) {$|\psi_i\rangle$ -- Quantum state};
\end{tikzpicture}
\caption{The Bloch sphere representation of a qubit, highlighting the basis states and angles defining the state vector's orientation.}
\label{fig:BlochSphere}
\end{figure}
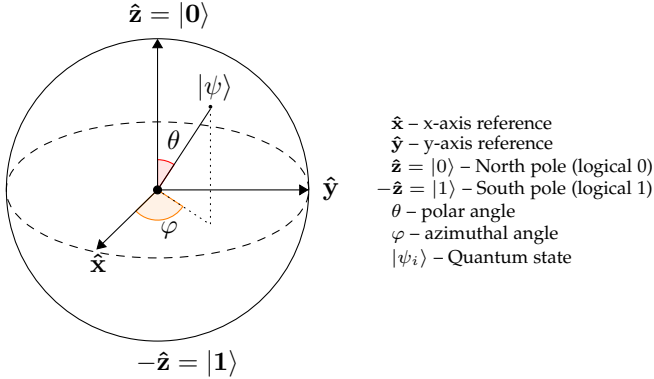

QC fundamentally exploits the non-classical phenomena of superposition, interference, and entanglement to encode and process information in ways that defy classical intuition~\cite{biamonte2017quantum,QFL6}. As illustrated in Fig. \ref{fig:BlochSphere}, in these systems, an input quantum state \( | \psi_i \rangle \) can exist in a superposition of basis states, enabling multiple computational pathways simultaneously, while interference modulates the probability amplitudes, enhancing desired outcomes and canceling others. Entanglement, a uniquely quantum phenomenon, correlates the states of spatially separated particles such that the overall quantum information \( \phi_i \) becomes inseparable \cite{preskill2012quantum}. Table \ref{tab:quantum_states} exemplifies canonical entangled states, each illustrating different facets of quantum correlations and their critical role in quantum algorithms.

\begin{table*}[h]
\centering
\caption{Key quantum states in QC.}
\begin{tabular}{|c|c|c|}
\hline
\textbf{Name}              & \textbf{Definition}                                                                                                                                                                                                                                                     & \textbf{Properties}                                                                                                                                                                                                                                       \\ \hline
Computational Basis States & \( |0\rangle \) and \( |1\rangle \)                                                                                                                                                                                                                                     & \begin{tabular}[c]{@{}c@{}}Form an orthonormal basis for a single qubit; \\ eigenstates of the Pauli-\(Z\) operator; \\ fundamental for digital quantum information\end{tabular}                                                                          \\ \hline
Superposition States       & \( |+\rangle = \frac{1}{\sqrt{2}}(|0\rangle + |1\rangle) \) and \( |-\rangle = \frac{1}{\sqrt{2}}(|0\rangle - |1\rangle) \)                                                                                                                                             & \begin{tabular}[c]{@{}c@{}}Exhibit maximal uncertainty in the computational\\  basis; eigenstates of the Pauli-\(X\) operator; \\ enable quantum parallelism\end{tabular}                                                                                 \\ \hline
Bell States                & \begin{tabular}[c]{@{}c@{}}$\Phi^+ = \frac{1}{\sqrt{2}}(|00\rangle + |11\rangle),\ \Phi^- = \frac{1}{\sqrt{2}}(|00\rangle - |11\rangle)$, \\ $\Psi^+ = \frac{1}{\sqrt{2}}(|01\rangle + |10\rangle),\ \Psi^- = \frac{1}{\sqrt{2}}(|01\rangle - |10\rangle)$\end{tabular} & \begin{tabular}[c]{@{}c@{}}Maximally entangled two-qubit states; essential\\  for protocols like quantum teleportation \\ and superdense coding; violate Bell inequalities\end{tabular}                                                                   \\ \hline
GHZ State                  & \( \frac{1}{\sqrt{2}}(|000\rangle + |111\rangle) \)                                                                                                                                                                                                                     & \begin{tabular}[c]{@{}c@{}}Exemplifies genuine multipartite entanglement; \\ highly sensitive to decoherence; useful in\\  quantum error correction and quantum secret sharing\end{tabular}                                                               \\ \hline
W State                    & \( \frac{1}{\sqrt{3}}(|001\rangle + |010\rangle + |100\rangle) \)                                                                                                                                                                                                       & \begin{tabular}[c]{@{}c@{}}Displays robust entanglement; remains partially \\ entangled even if one qubit is lost; \\ distinct from GHZ under LOCC transformations\end{tabular}                                                                           \\ \hline
Graph States               & \( |G\rangle = \prod_{(i,j) \in E} \text{CZ}_{ij} \, |+\rangle^{\otimes n} \)                                                                                                                                                                                           & \begin{tabular}[c]{@{}c@{}}Generalize cluster states; characterized by a graph\\  with vertices (qubits) and edges (entanglement via \\ controlled-\(Z\) gates); foundational for quantum \\ error correction codes and stabilizer formalism\end{tabular} \\ \hline
Pauli Gates &
\begin{tabular}[c]{@{}l@{}}
$\quad R_x(\theta)|0\rangle = \cos\frac{\theta}{2}|0\rangle - i\sin\frac{\theta}{2}|1\rangle$ \\[1ex]
$\quad R_y(\theta)|0\rangle = \cos\frac{\theta}{2}|0\rangle - \sin\frac{\theta}{2}|1\rangle$ \\[1ex]
$\quad R_z(\theta)|0\rangle = e^{-i\theta/2}|0\rangle$ \\
$\quad R_z(\theta)|1\rangle = e^{i\theta/2}|1\rangle$
\end{tabular}  &
\begin{tabular}[c]{@{}c@{}}Preserve quantum state norm and implement\\rotations about the x, y, and z axes enabling arbitrary\\single-qubit state manipulation; fundamental in\\constructing and decomposing complex quantum \\circuits and error correction schemes.\end{tabular} \\ \hline
  
\end{tabular}
\label{tab:quantum_states}
\end{table*}

Building on these foundational principles, Variational Quantum Circuits (VQC) \cite{biswas2024sensor} and Variational Quantum Algorithms (VQA) \cite{jin2024tetris} emerge as potent paradigms that iteratively optimize parameterized unitary operators, \( U(\theta) \). 
Here, \(U(\theta)\) denotes a sequenced quantum circuit comprising layers of single-qubit rotations \(R_{k}(\theta_{k})=e^{-i\theta_{k}\sigma_{k}/2}\) interleaved with entangling gates (for example, Controlled Not Gates (CNots), with the full operator taking the form  
\[
  U(\theta) \;=\; \prod_{\ell=1}^{L}\Bigl(\bigotimes_{q=1}^{Q}R_{q}^{(\ell)}(\theta_{q}^{(\ell)})\Bigr)\,E^{(\ell)}
\]  
where \(L\) is the number of layers, \(Q\) the qubit count, \(R_{q}^{(\ell)}\) the rotation on qubit \(q\) in layer \(\ell\), and \(E^{(\ell)}\) an entangling stage. They thus form smooth, differentiable mappings from a real parameter space into the unitary group \(U(2^n)\), preserving the norm of quantum states while enabling a rich family of rotations and entangling transformations \cite{Cerezo2021}.  
Their analytic dependence on \(\theta\) allows unbiased gradient evaluation via the parameter‐shift rule, making them amenable to variational optimization under NISQ limitations \cite{schuld2021effect}.  The circuit leverages this and transforms an input state \(\lvert \psi_{i}\rangle\) into an output \(\lvert \phi_{i}\rangle=U(\theta)\lvert \psi_{i}\rangle\), such that the measurement outcome \(
  y_{i}
  = \langle \phi_{i}\lvert M\rvert \phi_{i}\rangle
\) encodes the solution to a problem. This approach leverages quantum embeddings, which map classical data into high-dimensional \textit{Hilbert spaces} (a complete inner product space that serves as the framework for representing quantum states)\cite{kish2003quantum}, and quantum tensors that efficiently capture the intricacies of multi-qubit interactions~\cite{huggins2019towards}. Quantum Neural Networks (QNNs) adopt these variational frameworks, integrating an ansatz, which is a carefully constructed circuit template, into their architecture to learn complex representations~\cite{KAK1995259}. QAOA further exemplifies this strategy by embedding problem-specific constraints within its ansatz~\cite{mahroo2023learning}.

Yet, the promise of QC is tempered by practical challenges inherent to the NISQ era, where decoherence, gate imperfections, and qubit limitations impede flawless operation~\cite{chen2024nisq}. These challenges necessitate robust error mitigation strategies and innovative circuit designs to harness the true potential of quantum computation. In parallel, Fully Homomorphic Encryption (FHE) offers a classical countermeasure by enabling computations on encrypted data, thereby enhancing security in quantum communication protocols~\cite{dutta2024federated}. Quantum Key Distribution (QKD) further capitalizes on quantum mechanics to establish unconditionally secure channels, ensuring that cryptographic keys remain impervious to eavesdropping~\cite{yang2023survey}.   
As research continues to unravel the complexities of quantum tensors and refine variational techniques, the field strides steadily towards overcoming the limitations imposed by NISQ hardware.

\subsection{Overview of QFL}
The QFL framework, as illustrated in Fig.~\ref{fig:qfl-framework}, consists of two primary components:
\begin{enumerate}
    \item \textbf{Client:} To locally train the model using classical computational resources or quantum variational circuits \( U(\theta) \), or both in the case of a hybrid client. 
    \item \textbf{Central server:} To aggregate the classical or quantum parameters \( \theta^{(t)} \) and distribute the updated global parameters \( \theta^{(t+1)} \).
\end{enumerate}

Let the parameters of the quantum model in round \( t \) be denoted by \( \theta^{(t)} \). Each quantum client processes its local quantum data and updates the model. The flow of quantum information can be described as follows:
\begin{equation}
    \phi_i = U(\theta^{(t)}) | \psi_i \rangle
\end{equation}
where \( \phi_i \) is the quantum information of client i, $( U(\theta^{(t)})$ is the {\em parameterized unitary operator} for the quantum circuit, and \( | \psi_i \rangle \) denotes the {\em input quantum state} at the \( i \)-th node.\\

\subsubsection{At client}
\noindent Particularly, a quantum client performs the following:
\begin{enumerate}
    \item \textbf{Quantum circuit execution:} The unitary operation \( U(\theta) \) encodes the model parameters into the quantum circuit that evolves the input quantum state \( | \psi_i \rangle \) into the output state \( |\phi_i \rangle \) based on current parameters \( \theta_t \).
    \item \textbf{Measurement:} Quantum measurements \( M \) are performed on the evolved state \( |\phi_i \rangle \) to obtain observable outcomes:
    \begin{equation}
        y_i = \langle \phi_i | M | \phi_i \rangle
    \end{equation}
    where \( M \) is the measurement operator and \( y_i \) represents the measurement result corresponding to the local quantum training process.
    \item \textbf{Local gradient calculation:} Gradients are computed to update the parameter vector \( \theta^{(t)} \) at communication round \( t \). Typically, the parameter-shift rule is employed for VQC. For instance, when using a cross-entropy loss function for classification tasks, the gradient with respect to the \( j \)-th component, denoted as \( \theta^{(t)}_j \), is computed as
\begin{equation}
    \frac{\partial L}{\partial \theta^{(t)}_j} = L\!\Big(\theta^{(t)}_j + \frac{\pi}{2}\Big) - L\!\Big(\theta^{(t)}_j - \frac{\pi}{2}\Big)
\end{equation}
where \( \pi \) is the constant shift that leverages the periodicity of the parameterized quantum gates \cite{biamonte2017quantum}, and \( L \) denotes the loss function, such as the cross-entropy loss.
  
\end{enumerate}

\subsubsection{At server}
The computed local parameter updates \( \Delta \theta_i^{(t)} \) of $i^{th}$ client, for all $i$, are sent to the server for aggregation. Generally, the server then aggregates the local parameter updates \( \Delta \theta_i^{(t)} \) from \( N \) quantum clients using a weighted sum:
\begin{equation}
    \theta^{(t+1)} = \theta^{(t)} + \eta \sum_{i=1}^N w_i \Delta \theta_i^{(t)},
    \label{eqn:qfedavg}
\end{equation}
where \( \eta \) is the learning rate and \( w_i \) is the weight assigned to the \( i \)-th client based on its contribution. 
The updated global parameters \( \theta^{(t+1)} \) are distributed back to the quantum clients for the next iteration. Eq.~\ref{eqn:qfedavg} resembles the classical model aggregation algorithm, known as FedAvg~\cite{mcmahan2017communication}.

\subsection{Advantages over FL}
\begin{figure}[h]
    \centering
    \includegraphics[width=1\linewidth]{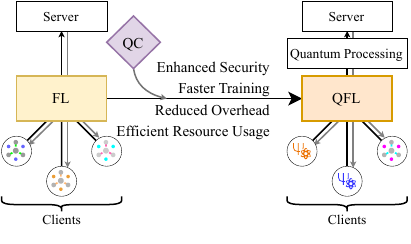}
    \caption{Advantages when FL meets QC.}
    \label{fig:FL-QFL}
\end{figure}

QFL merges the fields of QC and FL, using quantum technologies to enhance data processing, ML execution, and security. As depicted in Fig.~\ref{fig:FL-QFL}, the synergy of QC and the decentralized nature of FL addresses complex problems, improves learning, and overcomes FL limitations such as communication overhead and data privacy challenges. Traditional FL enables collaborative model training without sharing raw data, but struggles with communication costs and device limitations. QC integration improves processing power, thereby tackling FL issues. In particular, QFL offers the following advantages:

\begin{itemize}
    \item \textbf{Exponential speedup:} Quantum circuit such as \( U(\theta) \) enables faster computation of features and gradients, improving convergence for various ML tasks.
\item \textbf{Privacy preservation:} Quantum cryptographic protocols, such as quantum key exchange~\cite{yang2023survey} and quantum differential privacy~\cite{hirche2023quantum}, ensure secure transmission of parameters and quantum information.
    \item \textbf{Noise resilience:} Quantum measurements inherently introduce stochasticity, enhancing robustness against adversarial attacks on the model.

\end{itemize}
\new{
    Furthermore, Quantum-enabled protocols can curb both computation and communication costs in QFL beyond raw circuit acceleration. First, \emph{federated quantum embeddings} compress high-dimensional local features into amplitudes or low-qubit tensor-network states, which cuts per-round payload size and uplink time while preserving discriminative structure \cite{Qiao2025_Comst,QFL-TL-2,ballester_quantum_2025}. Second, \emph{parameter-sparse ansätze} constrain the trainable gate set and reduce forward–backward evaluations per local epoch, which shortens wall-clock time and lowers the number of parameters transmitted per round \cite{Qiao2025_Comst}. Third, \emph{post-quantum or quantum-secure aggregation} replaces heavyweight homomorphic encryption with lighter masking, which trims CPU cycles at clients and bytes on the wire \cite{QFL-SecAggPostQ1,QFL10,Zhang2025_SAReview}. Fourth, \emph{asynchronous aggregation} tolerates variable shot budgets and coherence windows, which limits idle waiting and reduces straggler amplification of both compute and message cost~\cite{Qiao2025_Comst,Chehimi2024_NSF}.
}

\begin{figure*}
    \centering
    \includegraphics[width=1\linewidth]{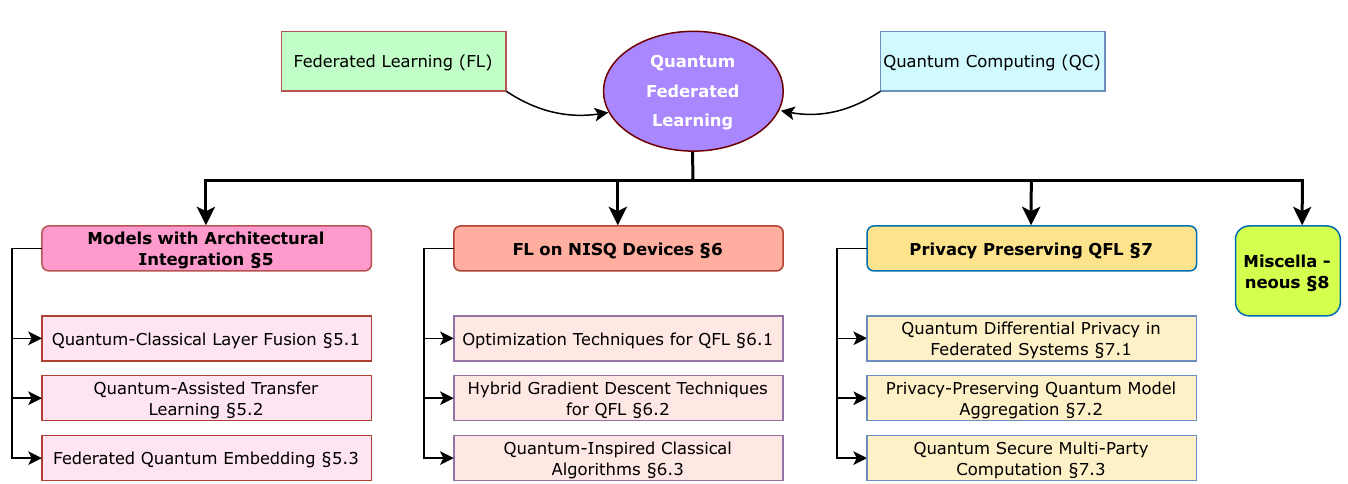}
    \caption{Illustration of the proposed taxonomy of the QFL research through a hierarchical framework.}
    \label{fig:taxonomy}
\end{figure*}

\section{Taxonomy of QFL}
\label{sec:Taxonomy}
This section provides a novel categorization of QFL, along with a quantitative and qualitative metric to systematically understand and compare QFL research.

\subsection{Novel Metrics for QFL Analysis}
The critical analysis of QFL requires new and novel dimensions that allow researchers to better understand recent works and innovate them to propose novel solutions. Thus, we introduce two crucial dimensions, namely \textit{Qubit Utilization Efficiency} ({\bf quantitative}) and \textit{Quantum Model Training Strategy} ({\bf qualitative}), to evaluate the practicality and performance of QFL frameworks. 

\subsubsection{Qubit Utilization Efficiency (QUE)} It focuses on the effective deployment of quantum resources, such as qubits and quantum gates, to achieve computational objectives within the constraints of NISQ devices. Let \(\mathrm{Q}_{alloc}\) be the total number of qubits allocated to the variational circuit, and let \(\mathrm{P}_{train}\) denote the trainable parameter count. If \(\mathrm{P}_{train}\) is not specified, we infer it as the product of the number of ansatz layers \(\mathrm{L}_{model}\) and the qubit count \(\mathrm{Q}_{alloc}\). This product captures two independent design choices, namely, the circuit depth (\(\mathrm{L}_{model}\)) and the hardware budget (\(\mathrm{Q}_{alloc}\)). Increasing either factor linearly raises the number of free parameters, which in turn enhances the variational ansatz’s representational power but also demands more quantum resources and classical overhead. For example, a three-layer hardware-efficient ansatz on six qubits yields \(\mathrm{P}_{train}=3\times6=18\). If \(\mathrm{L}_{model}\) is unreported, we fix a reference layer count \(\mathrm{L}_{model}=5\) since this is a common and sufficient layer count for most models. For any other missing values, suitable assumptions are made to estimate the required variables. 

\[
\mathrm{P}_{train} =
\begin{cases}
\mathrm{P}_{model},&\text{if specified}\\
\mathrm{L}_{model}\times \mathrm{Q}_{alloc},&\text{otherwise}\\
\end{cases}
\]

QUE is then penalized by factoring in entangling and any ancilla operations using the qubits assigned for them (\(\mathrm{Q}_{ancilla}\)). The QUE thus reads as
\begin{equation}
    \mathrm{QUE} = min\left \{1, \frac{\mathrm{P}_{train}}{\mathrm{L}_{model}(\mathrm{Q}_{alloc} + \mathrm{Q}_{ancilla} )} \right\}
    \label{eqn:QUE}
\end{equation}

 Values near unity indicate \textbf{High QUE}: almost every qubit participates in every layer, signaling compact circuit depth and minimal idle hardware. Intermediate values (\(0.5\le\mathrm{QUE}\le0.8\)) represent \textbf{Moderate QUE}: some qubits or layers remain unused, but the design remains reasonably dense. Lastly, values below 0.5 denote \textbf{Low QUE}: many qubit–layer slots sit idle, perhaps due to excessive ancilla or sparse gate placement. While no research in this survey falls under Low QUE currently, including the category would help in contrasting future QFL research on a comparable scale. For instance, the QCNN‐based QFL‐TL framework proposed in \cite{QFL-TL} uses a four‐qubit, four‐layer ansatz that achieves a QUE of 1.0, illustrating how shallow, well‐structured circuits can fully utilize qubit resources relative to gate overhead. On the other hand, the framework employed in \cite{FQE1}, an eight‐qubit, three‐layer QCNN ansatz, yields a QUE of approximately 0.80, reflecting moderate qubit–layer occupancy relative to its entangling overhead.

 This metric is particularly relevant for assessing scalability and carries particular weight on NISQ systems, where each idle qubit or under-utilized cycle magnifies decoherence risks and reduces practical throughput. 

\noindent $\bullet$ \textbf{Sensitivity Analysis}: Treating \(\mathrm{L}_{model}\) as a tunable parameter results in significant shifts in QUE across QFL implementations. In the QCNN‐TL model \cite{QFL-TL}, raising the layer count from 4 to 6 causes QUE to drop from 1.00 to 0.80. Meanwhile, the eight‐qubit embedding setup in \cite{FQE1} sees QUE decline from 0.80 at \(\mathrm{L}_{model}=3\) to 0.60 at \(\mathrm{L}_{model}=5\). Finally, in VQA‐based scheme described in \cite{huang2022quantum}, doubling \(\mathrm{L}_{model}\) from 4 to 8 reduces QUE from 0.75 to 0.50. These variations highlight a clear tension between circuit depth and qubit efficiency.

\noindent $\bullet$ \textbf{Security implications of scaling qubits and gates:} 
Beyond efficiency considerations, the implications of scaling qubit counts and circuit depth must also be examined from a security perspective. While QUE primarily captures the computational and communication overhead of scaling quantum circuits, an equally important dimension concerns the induced security risks. As the number of allocated qubits $Q_{\text{alloc}}$ and parameterized gates increases, the attack surface expands proportionally. 
The risk is accentuated in distributed QFL, where heterogeneous noise masks malicious parameter shifts and complicates attribution \cite{chu2025bvqcbackdoorstylewatermarkingscheme}. Recent studies in quantum network engineering highlight that deep NISQ circuits exacerbate vulnerability to manipulation due to the limited error-correction~\cite{khalid2024quantumnetwork}. Moreover, the integration of sustainable big AI frameworks with quantum broadcast channels has underscored how expanded gate-level programmability can inadvertently widen leakage channels if not secured by anonymous or semantic communication overlays~\cite{tariq2024semantic}. For QFL deployments, the implication is clear; efficiency gains from increasing $Q_{\text{alloc}}$ must be balanced with robust monitoring of parameterized gates, anomaly detection in PQC updates, and secure aggregation mechanisms that explicitly account for gate-level tampering. 

\subsubsection{Quantum Model Training Strategy (QMTS)}
This dimension captures the methodologies employed to train QFL models effectively. It distinguishes how clients integrate local quantum gradients into the global model.  Let \(\theta^{(t)}\) be the global parameter vector at round \(t\), and let \(\Delta\theta_{i}^{(t)}\) denote client \(i\)’s variational update computed on \(\mathrm{Q}_{alloc}\) qubits. The training strategies are broadly classified based on the following aggregation methods:

\noindent $\bullet$ \textbf{Synchronous aggregation:}  All \(K\) clients submit updates simultaneously. As stated in Eq.~\ref{eqn:qfedavg}, the server may perform a weighted average as:  
\[
  \theta^{(t+1)}
  = \theta^{(t)}
    + \eta \sum_{i=1}^{K}
      w_i\,
      \Delta\theta_{i}^{(t)}, \hspace{0.2cm} 
      where \hspace{0.2cm} 
      w_i=\frac{n_{i}}{\sum_{j}n_{j}}.
\]
This rule mirrors FedAvg where each \(\Delta\theta_{i}^{(t)}\) is derived from a parameter-shift quantum circuit \cite{QFL-type}. For example, \cite{QFL-QNGD} implements a synchronous aggregation scheme by collecting all local variational circuit parameter updates at each round and performing a single global update. This ensures that every client’s gradient contributes equally to the model synchronization and  mitigating drift. 

\noindent $\bullet$ \textbf{Asynchronous aggregation:}  Clients push updates at their own pace.  If client \(k\) sends at round \(t\) with staleness \(\tau_{k}\), then the update becomes:
\[
  \theta^{(t+1)}
  = \theta^{(t)}
    + \eta \sum_{i=1}^{K}
      w_i\,
      \Delta\theta_{i}^{(t-\tau_{k})}. 
\]
Stale corrections can accelerate slow nodes but may introduce bias.  A decay factor \(\gamma^{\tau_{k}}\) can thus temper outdated gradients. This allows for staggered participation to accommodate hardware variability, which is especially important in the NISQ devices. For example, \cite{QFL-AsynchAgg} employs an asynchronous aggregation, where each client pushes its encrypted quantum‐gradient message to the server. The server integrates these incoming updates on the fly.

\noindent $\bullet$ \textbf{Hybrid aggregation with classical–quantum fusion:} Let classical clients \(\mathcal{C}\) and quantum clients \(\mathcal{Q}\) each compute local updates.  The server then simply blends them as:  
\[
  \theta^{(t+1)}
  = \alpha\,\Bigl[\theta^{(t)} 
    + \eta\,\sum_{i\in\mathcal{C}}
      w_{i}\Delta\theta_{i}\Bigr]
  + (1-\alpha)\,\Bigl[\theta^{(t)} 
    + \eta\,\sum_{j\in\mathcal{Q}}
      w_{j}\Delta\theta_{j}\Bigr],
\]
where \(0\le\alpha\le1\) is used to balance the two modalities.  This strategy leverages classical expressivity alongside quantum expressivity by combining quantum-classical processes to leverage the strengths of both paradigms within a unified framework. The work~\cite{FQE2} exemplifies a hybrid classical–quantum fusion approach. Classical clients submit standard gradient averages, while quantum clients send variational circuit parameter deltas derived from their local quantum data, and then the server fuses these two modalities into a unified model update. 

By formalizing these three aggregation rules, the metric captures both the aggregation rhythm and the quantum‐classical interaction. Researchers can classify any QFL scheme by matching its update equation to either category.
The importance of this metric lies mainly in its ability to reveal the adaptability of QFL frameworks in heterogeneous and dynamic environments, such as those involving uncertainty of client availability or hardware reliability.

\subsection{Proposed Taxonomy}
Based on the main components, state-of-the-art techniques, and challenges, we propose a taxonomy dividing QFL research into three broad groups, as illustrated in Fig.~\ref{fig:taxonomy}: (i) models with architectural integration, (ii) FL on NISQ devices, and (iii) privacy-preserving QFL. This taxonomy aims to offer a structured and comprehensive view of existing work around key elements of the field. The structured perspective enables a deeper understanding of the QFL potential, guiding researchers in addressing the scalability, robustness, and efficiency challenges that define this rapidly evolving field. The proposed taxonomy not only highlights the strengths and limitations of current approaches but also identifies critical areas for future research.

\subsubsection{Models with Architectural Integration}
Architectural integration forms the backbone of QFL, defining how quantum and classical components collaborate to achieve distributed learning objectives. Hybrid architectures~\cite{Fusion1}, such as those that utilize VQCs or tensor-based networks, emphasize communication efficiency and scalability. These models often achieve significant reductions in communication overhead while maintaining high performance on specific tasks~\cite{ferrari2021compiler}. However, challenges persist in generalizing these architectures to heterogeneous and dynamic environments, particularly when addressing the trade-offs between communication optimization and model accuracy. Previous works highlight the need for robust architectural frameworks that balance computational demands with adaptability to real-world conditions~\cite{hisamori2024hybrid}.

\subsubsection{FL on NISQ Devices}
The integration of FL on NISQ devices represents a critical dimension, focusing on the practical deployment of FL within the constraints of quantum hardware. Models designed for NISQ devices often prioritize noise resilience and resource optimization, with techniques such as shallow circuits and error mitigation strategies playing a pivotal role. While prior surveys~\cite{narottama2020quantum, huang2022quantum, ren2023towards, gurung2023quantum, chehimi2024foundations, quy2024from, QFL-type, javeed2023quantum, quy2024from, saha2024multifaceted} have elaborated on performance metrics, including accuracy rates in noise-prone environments, the scalability of such systems remains a pressing challenge. For many quantum systems, additional qubits are used for error correction. While early-stage techniques like repetition codes and surface codes can partially reduce noise in NISQ systems, their scaling can present significant challenges in FL environments, which are detailed in Section~\ref{sec:FL on NISQ Devices}. 
This category analyzes the importance of developing hardware-aware algorithms that can accommodate the variability and limitations of NISQ devices.

\subsubsection{Privacy-Preserving QFL}
Privacy preservation is a cornerstone of QFL, addressing the dual objectives of securing sensitive data and ensuring robust federated training. Techniques such as quantum-enhanced differential privacy and homomorphic encryption have emerged as promising solutions~\cite{ullah2024quantum, broadbent2015quantum}, achieving strong privacy guarantees while maintaining competitive model performance. However, the computational overhead and trade-offs associated with these methods, particularly in noisy quantum environments, pose significant challenges.

\section{Models with Architectural Integration}
\label{sec:Models with Architectural Integration}


In QFL, models with architectural integration~\cite{QFL-QFuzzyNN,QFL-QLSTM,QFL-Tensor,QFL-Tensor2,QFL-QAOA,FQE2,QFL-QNGD,QFL-TL} represent a critical nexus between QC and distributed ML. These models aim to synergize quantum-enhanced computation with classical frameworks, promising unprecedented gains. 
The crux lies in integration, where the models leverage quantum advantages without falling prey to hardware limitations. This demands a delicate balance, optimizing theoretical quantum benefits against practical constraints while maintaining the core principles of data sovereignty and collaborative learning. Some common and useful terms are defined below:

\noindent $\bullet$ \textbf{VQC gradients:} VQCs are parameterized quantum circuits where gradients guide the update of circuit parameters to optimize performance. Their gradients are computed using the parameter-shift rule as:
    \[
    \frac{\partial f(\theta)}{\partial \theta_i} = f\left(\theta + \frac{\pi}{2} e_i\right) - f\left(\theta - \frac{\pi}{2} e_i\right)
    \]
where \(\pi\) represents the constant shift that leverages the periodic properties of the quantum gates, and \(e_i\) denotes the \(i\)-th standard basis vector, ensuring that the perturbation affects only the \(i\)-th parameter in \(\theta\).

\noindent $\bullet$ \textbf{Quantum Tensor Train Decomposition:} To quantify the compactness and efficiency of data representation between clients, high-dimensional data is decomposed as:
    \[
    X = \prod_{k=1}^d G_k,
    \]
    where \(X\) denotes the high-dimensional data tensor, and \(G_k\) represents the low-rank core tensors for each mode \(k\). 

\noindent $\bullet$ \textbf{Challenges:} QFL models offer theoretical advantages, but architectural integration is challenging. A major issue is noise and decoherence in quantum devices, which reduces computational accuracy and stability~\cite{shnirman2002noise}. VQCs, key to hybrid models, often face \textit{barren plateaus}, limiting scalability in deep quantum networks~\cite{mcclean2018barren}. The probabilistic nature of quantum measurements adds variability to gradient updates, complicating optimization and convergence in federated training~\cite{QFL-QNGD}. These challenges hinder the alignment of the quantum and classical frameworks~\cite{FQE2}.

Communication bottlenecks, especially in federated quantum settings where resources are scarce, hinder progress. Quantum-assisted models require frequent parameter exchanges through noisy channels, increasing quantum state transmission costs~\cite{streltsov2012quantum}. Integrating quantum tensor decompositions and hybrid layer fusion demands optimized circuit depths to avoid decoherence and high resource use~\cite{gelss2022low}. While tensor-network circuits and hierarchical updates can reduce complexity, their use is limited by the shortage of fault-tolerant qubits~\cite{anshu2024circuit}.

\subsection{Quantum-Classical Layer Fusion}
\begin{figure}[h]
    \centering
    \includegraphics[width=1\linewidth]{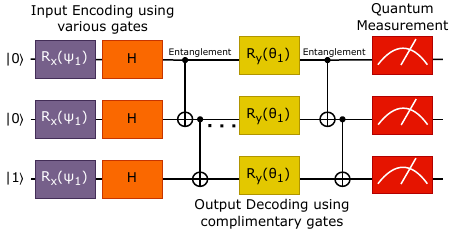}
    \caption{An example of a basic QNN comprising input encoding, entanglement with learnable parameters, output decoding to obtain new quantum states, and quantum measurement to obtain the calculated data.}
    \label{fig:QNN}
\end{figure}

Quantum-classical layer fusion~\cite{chen2024research, domingo2024hybrid} constitutes a fundamental architectural paradigm in hybrid learning models. This approach ingeniously replaces specific neural network layers, particularly those demanding substantial computational resources, with quantum equivalents. The quantum layers exploit the unique phenomena of quantum mechanics, superposition, and entanglement, to execute more intricate transformations on the input data. Concurrently, classical layers manage less resource-intensive operations, ensuring the model's viability for real-world applications. This synergistic integration of quantum and classical elements yields a hybrid neural network that, in theory, accelerates learning and enhances model accuracy~\cite{Fusion1}. QNNs (shown in Fig. \ref{fig:QNN}) are often employed in such architectures, leading to lightweight FL models \cite{park2025quantum}. This fusion architecture thus represents a critical step towards harnessing quantum computational advantages while navigating the constraints of current quantum hardware.

However, in practical deployments, fusion is not a trivial layer substitution but rather the careful placement of quantum modules at strategic points in the network. Each fusion point dictates the volume of quantum-classical communication, which, in a federated setting, directly affects aggregation latency and susceptibility to decoherence during data exchange. NISQ-era limitations remain a decisive factor: a restricted qubit allocation $Q_{\text{alloc}}$ constrains expressivity, deeper circuits amplify decoherence, and gate infidelity injects noise into the gradient update process. 
Moreover, bidirectional transfers between quantum processors and classical nodes increase round-trip latency.

\noindent \textbf{Potential applications:} Fusion-based architectures have already been adopted in several domains. In wearable healthcare analytics, quantum recurrent layers such as quantum LSTM have been integrated to improve energy expenditure estimation under constrained IoT device resources~\cite{tran2025quantumlstm}. In integrated satellite–ground remote sensing, quantum-enhanced convolutional layers can distill multi-modal features from heterogeneous sensors, thereby reducing the classical bandwidth requirement for model synchronization~\cite{khalid2025quantumfusion}. Privacy-sensitive NLP tasks can benefit from quantum embedding layers fused with classical attention mechanisms, enabling richer contextual encodings that are inherently more resistant to gradient inversion attacks.

\begin{figure}[h]
    \centering
    \includegraphics[width=1\linewidth]{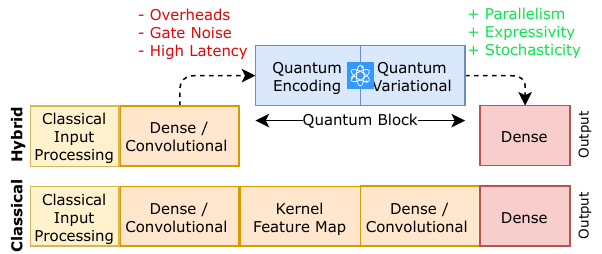}
    \caption{Illustrating quantum and classical layer fusion with its limitations (red-colored) and advantages (green-colored).}
    \label{fig:classical-quantum-fusion}
\end{figure}

Suppose a classical feature vector $\mathbf{h}_c$ is encoded into a quantum state $\ket{\psi(\mathbf{h}_c)}$. A parameterized quantum circuit $U(\boldsymbol{\theta})$ then transforms this state, and measurement results are obtained as
\begin{equation}
    {\mathbf{h}_q = \langle 0^{\otimes Q_{\text{alloc}}} | U^{\dagger}(\boldsymbol{\theta}) \, \hat{O} \, U(\boldsymbol{\theta}) | 0^{\otimes Q_{\text{alloc}}} \rangle,}
\end{equation}
where $\hat{O}$ is the measurement observable. The resulting $\mathbf{h}_q$ is concatenated with classical activations and propagated through subsequent layers in the FL pipeline, ensuring that quantum-derived features influence the global aggregation. Fig.~\ref{fig:classical-quantum-fusion} illustrates this architecture, where PQCs replace targeted classical layers and exchange measurement statistics with the classical stack during training.

\begin{table*}[h]
\caption{Critical analysis of the models with architectural integration.}
\centering
\begin{tabular}{|c|p{4cm}|p{4cm}|p{3.5cm}|p{3.5cm}|}
\hline
\textbf{Paper} & \textbf{Strengths} & \textbf{Limitations} & \textbf{QUE [Major reason]} & \textbf{QMTS}\\
\hline
\cite{QFL-TL} 
  & Effective use of QCNNs for scalable learning; first implementation of quantum weights 
  & High noise sensitivity and limited real-world privacy-preservation 
  & \begin{tabular}[t]{@{}l@{}}High\\[0.3ex]{[}shallow circuit design{]}\end{tabular} 
  & Asynchronous aggregation \\
\hline
\cite{FQE1}
  & Pure QFL framework with quantum data, utilizing QCNN 
  & Potential scalability issues; vulnerable to attacks 
  & \begin{tabular}[t]{@{}l@{}}Moderate\\[0.3ex]{[}8-qubit QCNN circuits{]}\end{tabular} 
  & Asynchronous aggregation \\
\hline
\cite{QFL-QNGD} 
  & High efficiency in training and reduced communication rounds 
  & Poor robustness with non-uniform and complex data distributions 
  & \begin{tabular}[t]{@{}l@{}}High\\[0.3ex]{[}variational circuit\\design{]}\end{tabular} 
  & Synchronous aggregation \\
\hline
\cite{FQE2} 
  & Comprehensive comparative analysis across FL frameworks; includes quantum network security insights 
  & Lacks practical benchmarks or real deployment results 
  & \begin{tabular}[t]{@{}l@{}}Moderate\\[0.3ex]{[}PQC-based circuits{]}\end{tabular} 
  & Hybrid aggregation \\
\hline
\cite{QFL-QAOA} 
  & Strong privacy mechanisms for non-IID data; adaptable to distributed systems 
  & High communication overhead and potential scalability issues 
  & \begin{tabular}[t]{@{}l@{}}High\\[0.3ex]{[}efficient VQA circuits{]}\end{tabular} 
  & Asynchronous aggregation \\
\hline
\cite{QFL-Tensor2} 
  & Efficient handling of communication and computational cost 
  & Limited adaptability in highly distributed or large-scale settings 
  & \begin{tabular}[t]{@{}l@{}}Moderate\\[0.3ex]{[}tensor decomposition\\strategies{]}\end{tabular} 
  & Synchronous aggregation \\
\hline
\cite{QFL-Tensor} 
  & High data privacy integration; effective federated updates 
  & Complex implementation may limit practical scalability 
  & \begin{tabular}[t]{@{}l@{}}Moderate\\[0.3ex]{[}hierarchical tensor\\strategies{]}\end{tabular} 
  & Hybrid aggregation \\
\hline
\cite{QFL-QLSTM} 
  & Efficient for time-series data; reduced training overhead 
  & Low adaptability to non-time-series data and basic noise handling 
  & \begin{tabular}[t]{@{}l@{}}High\\[0.3ex]{[}optimized QLSTM\\structure{]}\end{tabular} 
  & Asynchronous aggregation \\
\hline
\cite{QFL-QFuzzyNN} 
  & Robust noise resilience; adaptive training using fuzzy logic 
  & High complexity in setup and limited broad dataset testing 
  & \begin{tabular}[t]{@{}l@{}}Moderate\\[0.3ex]{[}adaptive QFNNs{]}\end{tabular} 
  & Asynchronous aggregation \\
\hline
\end{tabular}
\label{tab:architectural_critical_analysis}
\end{table*}

One key advantage of layer fusion is that it does not require fully developed QNNs. This is beneficial as this fusion may prove to be infeasible due to the limitations of existing quantum hardware. Instead, by partitioning the neural network, only specific layers leverage quantum resources, making this a more practical and scalable solution. An alternative solution involves the implementation of a weighted personalization algorithm that addresses data heterogeneity and model drift~\cite{gurung2025performance}. Recent studies demonstrated that hybrid models that use this technique can outperform fully classical models in specific tasks such as quantum chemistry simulations and optimization problems~\cite{Chem1, optimization1}. Despite its promise, this model is still heavily constrained by the need for efficient quantum gates and error-correcting mechanisms that are not yet mature~\cite{error1}.

An examination of the QUE values in Table~\ref{tab:architectural_critical_analysis} shows that most layer-fused models operate with moderate to high qubit‐layer occupancy (typically in the range of 67\% to 100\%). Shallow QCNN hybrids like \cite{QFL-TL} and \cite{FQE1} exploit asynchronous updates to capitalize on their high QUE, since clients can independently push low-latency gradient estimates without waiting for global synchronization. In contrast, approaches with deeper variational circuits, such as \cite{QFL-QNGD} and \cite{QFL-Tensor2}, favor synchronous aggregation to ensure that their more complex parameter landscapes converge cohesively. Despite these strengths, layer fusion introduces new challenges. The heterogeneity of classical and quantum execution pipelines can cause unpredictable latency spikes.

\subsection{Quantum-Assisted Transfer Learning}
Quantum-assisted transfer learning~\cite{QFL-TL,QFL-TL-2} introduces the potential to enhance classical Transfer Learning (TL) techniques by leveraging quantum computing for feature extraction and optimization. This approach enables better generalization across tasks using quantum algorithms to identify and transfer rich feature representations.

A notable research in~\cite{QFL-TL} explored \textit{Federated Quantum ML}, utilizing hybrid quantum-classical models. By combining quantum circuits with pre-trained classical models, this work accelerates convergence without sacrificing performance, significantly addressing the scalability challenge in NISQ devices~\cite{Preskill2018}. The proposed methodology is also the first to implement quantum weights, enabling a fully quantum ring network topology for QFL. 
Another work~\cite{QFL-TL-2} proposed a novel TL architecture, which leverages quantum networks and pre-trained classical deep neural networks, to improve the adaptation of the model between related tasks. 
A critical insight from this study is that quantum states can enhance the capacity for information transfer, particularly when dealing with high-dimensional data. 

An assessment of \(\mathrm{QUE}\) for the transfer‐learning schemes suggests moderate to high utilization of qubit–layer resources.  In \cite{QFL-TL}, shallow QCNN circuits (e.g.\ \(\mathrm{Q}_{alloc}=4\), \(L=4\)) yield \(\mathrm{QUE}\approx1.0\), which pairs well with its asynchronous update strategy to minimize idle time on quantum nodes.  By contrast, the hybrid ring architecture in \cite{QFL-TL-2} employs slightly deeper VQCs (e.g.\ \(\mathrm{Q}_{alloc}=4\), \(L=5\)), giving \(\mathrm{QUE}\approx0.80\); this design favors synchronous aggregation so that rich quantum features are consistently merged with classical weights in each round.

\subsection{Federated Quantum Embedding}
Federated Quantum Embedding (FQE)~\cite{FQE1,FQE2} represents a cutting-edge approach in which classical data are encoded into quantum states, allowing richer and more expressive feature representations across FL systems. This method enables quantum devices to extract complex features from classical datasets by leveraging the quantum states, which can also be shared and aggregated in an FL framework.

The work in~\cite{FQE1} proposed a framework for processing purely quantum data in QFL, demonstrating that quantum embeddings can significantly reduce the communication cost associated with distributed learning. 
The key advantage of FQE lies in its ability to compress high-dimensional classical data into compact quantum states by employing techniques such as {\em quantum tensor train decomposition} \cite{bhatia2025application}, thus optimizing both communication efficiency and learning performance. The proposed model implements the \textit{Quantum Convolutional Neural Networks} (QCNN)-QFL framework using the first federated quantum dataset. However, the challenge remains to manage the quantum noise introduced during the embedding process, which can degrade the quality of the shared quantum states, thus necessitating robust error mitigation techniques to ensure stable model updates. 
Building on this, another study~\cite{FQE2} investigated FQE in classical and quantum networks, emphasizing its potential to boost learning in federated settings with large-scale distributed data. It examines efficient quantum embedding aggregation and tackles coherence loss during data transmission. Furthermore, novel quantum communication protocols are suggested to maintain data integrity in noisy channels, offering a promising solution to the challenge of noisy quantum channels \cite{chu2023cryptoqfl}. 
By encoding sensitive data into quantum states, FQE is inherently harder for adversaries to reverse-engineer the original information from the embeddings, thus enhancing the privacy guarantees.

An evaluation of \(\mathrm{QUE}\) for the embedding‐based models shows that the QCNN‐QFL (\(\mathrm{Q}_{alloc}=4\), \(L=3\)) framework of \cite{FQE1} achieves \(\mathrm{QUE}=0.80\), reflecting its compact three‐layer encoding circuit and justifying its asynchronous update strategy, which minimizes idle qubit time during on‐demand embedding. In contrast, the comparative study of \cite{FQE2} across classical and quantum networks uses a slightly deeper ansatz (\(\mathrm{Q}_{alloc}=6\), \(L=4\)), yielding \(\mathrm{QUE}=0.75\), and favors a hybrid classical–quantum fusion approach to merge classical gradient averages with quantum embedding deltas. Apart from embedding noise and coherence loss, these schemes confront the overhead of repeated state preparation and measurement for each data batch, which can dominate communication cost as dataset sizes grow. 

\noindent \textbf{Key insights (Models with Architectural Integration):} From the literature on architectural integration, we observe the following points: 
\begin{itemize}
    \item Models with layer fusion split neural networks into quantum and classical layers. The quantum layers tackle complex tasks, leveraging superposition and entanglement, and classical layers handle simpler operations, ensuring scalability. Mitigating quantum noise and optimizing the circuit depth in NISQ devices are major challenges yet to be addressed. 
    
    \item Through TL, quantum circuits can be used for feature extraction and improving model generalization in FL tasks. This approach can speed up convergence and enhance performance on non-IID datasets, but is limited by communication overhead.

    \item Models with FQE improve feature representation and reduce communication overhead in federated networks. Using the high-dimensionality of quantum states, QFE enhances accuracy and privacy but requires strong error correction to manage noise.
 
\end{itemize}
\noindent Finally,  to provide a comprehensive overview, Table~\ref{tab:architectural_critical_analysis} presents the models that incorporate architectural integration, highlighting their respective strengths, limitations, QUE, and training strategies.

\section{FL on NISQ Devices}
\label{sec:FL on NISQ Devices}
\begin{figure}[h]
    \centering
    \includegraphics[width=1\linewidth]{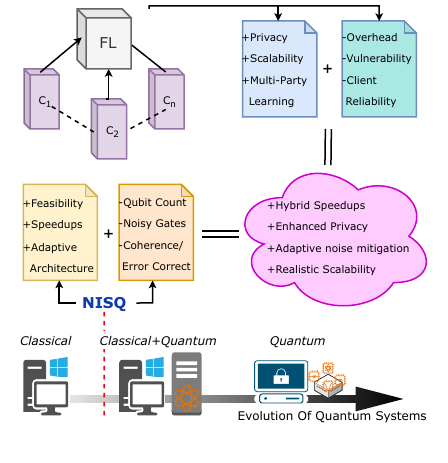}
    \caption{Pros and cons of NISQ and FL systems individually, as well as the benefits obtained when combining both into a single comprehensive framework. `+' and `-' represent pros and cons, respectively.}
    \label{fig:FL-NISQ}
\end{figure}

FL on NISQ devices is an emerging field in which quantum computational power is harnessed to improve the scalability and efficiency of decentralized ML models. The evolution of quantum systems has led to the NISQ era; Fig.~\ref{fig:FL-NISQ} depicts the pros and cons when FL is deployed on NISQ. Despite its potential, the integration of NISQ devices into FL systems presents significant challenges because of the limitations of current quantum hardware. This section explores such challenges under optimization techniques for QFL, hybrid gradient descent methods, and quantum-inspired classical algorithms, attempting to bridge the gap between classical and quantum computation.

\noindent $\bullet$ \textbf{QAOA ansatz:} Central to QAOA is the variational ansatz alternating between cost and mixing Hamiltonians:
    \[
    U(C, \gamma) U(B, \beta) |\psi\rangle
    \]
    where \(U(C, \gamma)\) and \(U(B, \beta)\) are unitary operators parameterized by \(\gamma\) and \(\beta\), respectively, applied iteratively to approach the solution of combinatorial optimization problems.

\noindent $\bullet$ \textbf{Federated Multi-Objective Optimization (FedMGDA+):} To balance fairness and robustness, FedMGDA+ uses Pareto optimization to solve:
    \[
    \min_x \{F_1(x), F_2(x), \dots, F_m(x)\}
    \]
    subject to ensuring that no objective function is unduly sacrificed during optimization.

\noindent $\bullet$ \textbf{Challenges:}
While offering quantum computational advantages, NISQ devices are inherently constrained by high levels of noise~\cite{cao2021nisq, yetis2021investigation}, limited qubit count~\cite{tannu2019not,nash2020quantum}, and low gate fidelity~\cite{chen2023high, berrios2012high}. The primary challenge in utilizing these devices within an FL framework is managing quantum noise. Quantum states, especially in multi-qubit systems, are highly sensitive to decoherence, leading to loss of information and reduced model accuracy~\cite{Preskill2018}. Additionally, the relatively small number of qubits limits the complexity of quantum computations, restricting the depth of quantum circuits involved in federated tasks.

Most quantum systems use additional qubits for Quantum Error Correction (QEC) techniques. Some early-stage techniques, such as repetition codes and surface codes, provide partial mitigation of noise in NISQ systems~\cite{khalid2024quantumnetwork}. However, their scaling characteristics pose distinct challenges in FL. Each client node operating under QEC must allocate additional qubits for encoding logical qubits, increasing $Q_{\text{alloc}}$ well beyond the original model requirements. In large-scale FL deployments, this qubit overhead compounds across clients, limiting the feasible number of participating devices. 
In asynchronous settings, heterogeneous QEC performance across clients can lead to model inconsistency if updates are based on variably protected computations. These constraints imply that while early QEC can enhance noise resilience, its direct adoption in FL must be balanced against the cost in qubit resources, communication time, and aggregation stability.
It indicates that error mitigation techniques, rather than full QEC, may be the most viable route forward to realize FL on NISQ devices~\cite{ErrorMitigation}.

\new{
Furthermore in practical QFL deployments, clients rarely have identical logical-qubit capacities, coherence times, or noise characteristics, and a rigid assumption of a single uniform PQC architecture would either exclude weaker devices or inject unstable gradients that slow or destabilize convergence \cite{QFL-TL-2}. A more realistic strategy treats hardware variation as a structured design constraint i.e., the server specifies a nested family of ansatzes in which low-depth, low-qubit subcircuits constitute a shared core and deeper blocks extend this core for clients with larger logical budgets, analogous in spirit to model-heterogeneous FL, but adapted to parametrized quantum circuits \cite{diao_heterofl_2021}. Each client then instantiates the deepest sub-ansatz compatible with its logical-qubit and shot budget, updates only the parameters it supports, and contributes these to aggregation, while the coordinator averages each parameter block over the subset of clients that implements it~\cite{diao_heterofl_2021,han_quorus_2025}. This perspective aligns with recent QFL frameworks that employ layerwise or blockwise objectives to handle varying circuit depths, where heterogeneous-depth clients optimize different truncations of a shared PQC and still produce a coherent global model \cite{han_quorus_2025}. Extremely constrained clients can still play meaningful roles by hosting shallow encoders, measurement heads, or hybrid split-learning frontends that interface with quantum backends. This verifies that limited local quantum footprint can be combined with stronger remote resources while preserving data locality~\cite{cowlessur_hqsl_2025}. Combined with the temporal mechanisms, in which asynchronous or semi-synchronous aggregation assigns staleness and depth-aware weights to updates, these constructions ensure that devices with diverse logical-qubit resources contribute systematically rather than sporadically
}

\begin{table*}[htbp]
\caption{Critical analysis of prior research when FL deployed on NISQ devices.}
\centering
\begin{tabular}{|c|p{4cm}|p{4cm}|p{3.5cm}|p{3.5cm}|}
\hline
\textbf{Paper} & \textbf{Strengths} & \textbf{Limitations} & \textbf{QUE [Major reason]} & \textbf{QMTS} \\
\hline
\cite{QFL-Optimization} 
  & Demonstrates adaptive methods that improve convergence in nonconvex FL 
  & Sensitive to heterogeneity, client drift not fully resolved 
  & \begin{tabular}[t]{@{}l@{}}Moderate\\[0.3ex]{[}adaptive optimization{]}\end{tabular} 
  & Synchronous aggregation \\
\hline
\cite{QFL-MultiOptimization} 
  & Integrates Pareto optimization ensuring fairness and robustness 
  & Complexity in multi-objective trade-offs can slow convergence 
  & \begin{tabular}[t]{@{}l@{}}Moderate\\[0.3ex]{[}multi-objective balance{]}\end{tabular} 
  & Asynchronous aggregation \\
\hline
\cite{QFL-QNGD} 
  & Faster convergence and reduced communication rounds in QFL 
  & Limited generalization to complex datasets with significant noise 
  & \begin{tabular}[t]{@{}l@{}}High\\[0.3ex]{[}variational circuit{]}\end{tabular} 
  & Synchronous aggregation \\
\hline
\cite{QFL-QAOA} 
  & Comprehensive review of QAOA and its variants; hardware-focused insights 
  & No practical implementation insights for distributed systems 
  & \begin{tabular}[t]{@{}l@{}}High\\[0.3ex]{[}optimized circuit depth{]}\end{tabular} 
  & Hybrid aggregation \\
\hline
\cite{QFL-QLSTM} 
  & Effective for time-series data processing with reduced training time 
  & Limited robustness in non-time-series contexts 
  & \begin{tabular}[t]{@{}l@{}}High\\[0.3ex]{[}efficient QLSTM structure{]}\end{tabular} 
  & Asynchronous aggregation \\
\hline
\cite{QFL-TL-2} 
  & Proposes hybrid quantum-classical training; preserves data privacy 
  & Scalability issues for larger quantum data sets 
  & \begin{tabular}[t]{@{}l@{}}Moderate\\[0.3ex]{[}hybrid circuit integration{]}\end{tabular} 
  & Synchronous aggregation \\
\hline
\cite{QFL-QCNN} 
  & Integrates QCNNs for non-IID data in medical imaging; reduces communication rounds 
  & Scalability and noise resilience not extensively addressed 
  & \begin{tabular}[t]{@{}l@{}}High\\[0.3ex]{[}shallow quantum circuits{]}\end{tabular} 
  & Synchronous aggregation \\
\hline
\cite{QFL2} 
  & Adaptable across quantum hardware; strong results on multiple datasets 
  & Accuracy fluctuations due to noise and gate fidelity, limited large-scale insights 
  & \begin{tabular}[t]{@{}l@{}}Moderate\\[0.3ex]{[}adaptive qubit usage{]}\end{tabular} 
  & Asynchronous aggregation \\
\hline
\end{tabular}
\label{tab:quantum_federated_analysis}
\end{table*}

\subsection{Optimization Techniques for QFL}
To overcome the inherent limitations of NISQ devices, various optimization strategies have emerged. Central to this effort are VQAs, which operate by embedding quantum circuits within a classical optimization loop. In this framework, quantum circuits explore a vast solution space, while classical processors adjust variational parameters based on performance evaluations. The classical optimizer iteratively refines the parameters until a convergence criterion is met, ensuring that the workload is effectively shared between the quantum and classical components~\cite{QFL-Optimization}. Such a hybrid approach has been crucial for leveraging quantum advantages without overwhelming the capacity of NISQ devices. However, optimization in hybrid systems can be resource-intensive. Using ML, the authors in~\cite{ML-Optimization} attempted to predict optimal quantum parameters with QAOA. Remarkably, their technique reduces the number of iterations in the optimization loop by as much as 65.7\%. Instead of initiating each optimization loop from scratch, they harnessed correlations between lower-depth and higher-depth QAOA parameters. This intelligent initialization significantly reduces computational cost, accelerating convergence. 

Beyond classical ML, there is a growing focus on enhancing distributed quantum systems. QFL tackles optimization challenges in decentralized and heterogeneous environments. Techniques like FedMGDA+~\cite{QFL-MultiOptimization} are designed to ensure fairness and robustness, even when faced with adversarial conditions and uneven data distribution. The adaptability of these methods makes them particularly valuable for ensuring stability in real-world quantum networks. 
QFL researchers have shown that the transfer of parameters from simpler quantum tasks can speed up optimization for complex problems, reducing the overall computational burden~\cite{QFL-TL-2}. This cross-task learning is pivotal in accelerating multitask optimization within quantum systems.

An evaluation of $\mathrm{QUE}$ for optimization‐focused QFL methods shows that VQA‐based schemes like the adaptive optimizer in \cite{QFL-Optimization} typically employ moderate‐depth circuits (e.g.\ \(\mathrm{Q}_{alloc}=5\), \(L=4\)) yielding \(\mathrm{QUE}\approx1.0\), which aligns with their preference for synchronous global updates to exploit full‐batch curvature corrections. In contrast, Pareto‐based multi‐objective optimizers such as FedMGDA+~\cite{QFL-MultiOptimization} often run on slightly shallower ansatze (\(\mathrm{Q}_{alloc}=4\), \(L=3\)), giving \(\mathrm{QUE}\approx0.80\) and benefiting from asynchronous local updates that accommodate imbalanced data distributions. TL accelerators like the parameter‐reuse strategy in \cite{QFL-TL-2} sit between these extremes with \(\mathrm{QUE}\approx0.90\) and favor hybrid classical–quantum fusion to blend pretrained classical weights with fresh quantum gradients. Beyond the optimization benefits, these methods must address the overhead of metric‐tensor estimation in ~\cite{QFL-QNGD}, the staleness introduced by lock‐free updates in \cite{QFL-MultiOptimization}, and the calibration drift that arises when classical weights are merged with quantum‐tuned parameters in transfer scenarios.

\subsection{Hybrid Gradient Descent Techniques for QFL}
Gradient descent is at the heart of most ML algorithms, and it plays a pivotal role in updating parameters across distributed nodes in FL. As quantum devices become more integrated with FL systems, hybrid gradient descent techniques have emerged. These techniques leverage the computational power of quantum processors to compute partial gradients or assist in optimizing specific model parameters. The combination of quantum and classical resources accelerates convergence by efficiently partitioning the tasks between quantum and classical nodes~\cite{optimization2}.

A notable research in~\cite{QFL-QNGD} introduced Federated Quantum Natural Gradient Descent (FQNGD), a novel optimization method for QFL, utilizing VQCs to enhance convergence and reduce communication overhead. The experimental results on the MNIST dataset showed superior accuracy (more than 99\% for binary classification) over traditional SGD methods. This technique markedly enhances convergence rates, especially within the realm of NISQ devices~\cite{optimization3}. Furthermore, by allowing quantum devices to compute gradients for only a subset of parameters, hybrid systems can dynamically balance computational loads and reduce training bottleneck~\cite{optimization4}. However, reliance on block-diagonal approximations of the Fubini-Study could limit robustness, as it may overlook complex entanglements in deeper circuits, impacting practical performance. 

Analysis of $\mathrm{QUE}$ for hybrid gradient methods shows that FQNGD’s variational circuits (\(\mathrm{Q}_{alloc}=5\), \(L=4\)) achieve \(\mathrm{QUE}\approx1.0\), matching its preference for synchronous aggregation to exploit full‐batch curvature corrections via the quantum Fisher metric \cite{QFL-QNGD}. Asynchronous update schemes, exemplified by the Paillier‐secured protocol in \cite{QFL-AsynchAgg}, typically run on shallower ansatz (\(\mathrm{Q}_{alloc}=4\), \(L=3\)), yielding \(\mathrm{QUE}\approx0.80\) and benefiting from low-latency, lock‐free gradient pushes. Fusion strategies that combine classical FedAvg with quantum natural gradients, such as the hybrid model in \cite{FQE2} operate at moderate QUE (\(75\%\)–\(85\%\)) and balance robustness and acceleration by tuning a mixing coefficient. However, these methods still face the challenge of efficiently estimating and inverting high-dimensional metric tensors on noisy hardware.

\noindent $\bullet$ {\bf Hybrid gradient implementation:}
To optimally implement hybrid gradient descent in a QFL setting, it is important to design it such that it respects both the characteristics of PQCs and the practical constraints of distributed FL training. Let us partition the global parameter vector into quantum and classical components as 
\(
\boldsymbol{\theta} \;=\; (\boldsymbol{\theta}_{Q},\, \boldsymbol{\theta}_{C}),
\)
reflecting the fused architecture \cite{Mitarai2018QCL,Schuld2019AnalyticGradients}. This is because PQC parameters are updated through measurement-based estimators that differ fundamentally from the backpropagation-based updates applied to classical layers. Each client \(k\) then  evaluates its local objective \(\mathcal{L}_k\) and constructs a hybrid gradient that blends quantum and classical components:
\[
\nabla \mathcal{L}_k(\boldsymbol{\theta}) \;=\; \alpha\,  \nabla_{\boldsymbol{\theta}_{Q}} \mathcal{L}_k(\boldsymbol{\theta}) \;+\; (1-\alpha)\, \nabla_{\boldsymbol{\theta}_{C}} \mathcal{L}_k(\boldsymbol{\theta}),
\]

where the coefficient \(\alpha \in [0,1]\) regulates the relative contribution of quantum updates. This convex weighting provides flexibility, allowing the update dynamics to be tuned in response to hardware noise levels, QPU queue times, or task-specific sensitivity to quantum features. For the quantum component, gradients are estimated via the parameter-shift rule, which for a parameter \(\theta_{Q,i}\), evaluated at 2 shifted points $\pm s$ takes the form

\[
\frac{\partial \mathcal{L}_k}{\partial \theta_{Q,i}}
=\frac{1}{2}\!\left[
\mathcal{L}_k(\boldsymbol{\theta}_{Q} + s\,\mathbf{e}_i,\,\boldsymbol{\theta}_{C})
-
\mathcal{L}_k(\boldsymbol{\theta}_{Q} - s\,\mathbf{e}_i,\,\boldsymbol{\theta}_{C})
\right]
\]

with shift \(s=\pi/2\) and \(\mathbf{e}_i\) the \(i\)th standard basis vector \cite{Schuld2019AnalyticGradients,Crooks2019ParamShift}. This rule is compatible with a wide range of NISQ devices, avoids the overhead of full adjoint differentiation, and introduces minimal circuit depth change, thereby preserving coherence times. As discussed above, when applicable, the PQC gradient can be preconditioned using the quantum natural gradient. This involves estimating the Fubini–Study metric \(\mathbf{F}_k\) and rescaling the gradient via
\(
\widetilde{\nabla}_{\boldsymbol{\theta}_{Q}} \mathcal{L}_k \;=\; \mathbf{F}_k^{-1}\, \nabla_{\boldsymbol{\theta}_{Q}} \mathcal{L}_k,
\)
which aligns the update direction with the local geometry of the Hilbert space. This approach mitigates the effects of barren plateaus and poor conditioning, enabling faster and more stable convergence. The local update is then performed with learning rate \(\eta\) as
\(
\boldsymbol{\theta}^{\,t+1}_k
=
\boldsymbol{\theta}^{\,t}
-
\eta\,\nabla \mathcal{L}_k(\boldsymbol{\theta}^{\,t}),
\)
or using the preconditioned gradient in the quantum block if natural gradient correction is applied. Conducting multiple local steps before communication balances QPU access and reduces communication overhead in federated setups. Aggregation at the server can proceed synchronously, where sample-proportional weights \(p_k = n_k / \sum_{j\in\mathcal{S}_t} n_j\) yield
\(
\boldsymbol{\theta}^{\,t+1}
=
\boldsymbol{\theta}^{\,t}
- 
\eta \sum_{k\in\mathcal{S}_t} p_k\,\nabla \mathcal{L}_k(\boldsymbol{\theta}^{\,t}),
\)
ensuring that quantum and classical parameters evolve together across clients. Alternatively, in asynchronous settings, stale updates can be down-weighted via a decay factor \(w_k(\Delta t_k) \in (0,1]\) dependent on staleness \(\Delta t_k\):
\[
\boldsymbol{\theta}^{\,t+1}
=
\boldsymbol{\theta}^{\,t}
-
\eta \sum_{k\in\mathcal{A}_t}
p_k\, w_k(\Delta t_k)\,
\Big[
\alpha\, \widetilde{\nabla}_{\boldsymbol{\theta}_{Q}} \mathcal{L}_k
+
(1-\alpha)\, \nabla_{\boldsymbol{\theta}_{C}} \mathcal{L}_k
\Big].
\]

This correction limits the influence of outdated PQC updates while retaining the throughput benefits of asynchronous aggregation for classical components. The triplet \((\alpha,\eta,w_k)\), together with optional natural gradient preconditioning, forms the core control space for adapting the hybrid federated optimization loop to varying noise models, hardware availability, and task difficulty.

\subsection{Quantum-Inspired Classical Algorithms}

With the introduction of quantum neural computation by~\cite{QNC}, QNN received significant attention from the QFL community.
This pioneering theoretical framework is a crucial advancement in the integration of quantum computation concepts with traditional neural network models, setting the stage for subsequent studies of quantum-inspired classical algorithms. Building on this foundation, numerous quantum-enhanced classical techniques have been developed, each targeting particular computational constraints of classical systems while utilizing quantum benefits.

\begin{figure}[h]
    \centering
    \includegraphics[width=1\linewidth]{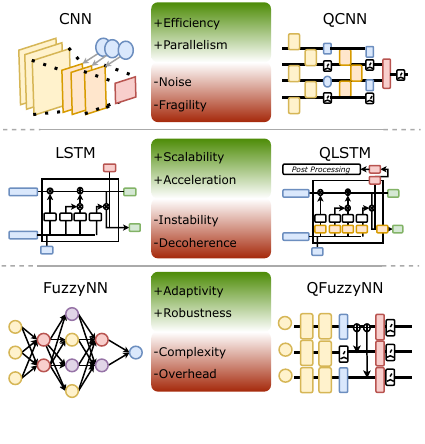}
    \caption{Advantages and disadvantages of classical neural network frameworks when combined with quantum computing. `+' and `-' represent advantages and disadvantages, respectively.}
    \label{fig:Quantum-Counterparts}
\end{figure}

Building on the challenges of quantum circuit stability, FedQNN~\cite{QFL2} presents a more generalized federated quantum framework for various datasets, including genomics and healthcare. 
A critical strength of the framework is its adaptability to different quantum hardware architectures, evidenced by evaluations on IBM QPUs. 
The authors in~\cite{QFL-QCNN} introduced a novel FL framework using QCNNs for privacy-preserving collaborative learning, particularly suited for non-IID medical data distributions. The novelty lies in the use of quantum circuits to enhance feature extraction and classification within a federated setting.
The development of the \textit{Quantum Long Short-Term Memory (QLSTM)}~\cite{QFL-QLSTM} network represents a significant advancement aimed at overcoming the challenges faced by classical LSTMs~\cite{sadhwani20245g}, especially in managing long-term dependencies in time-series data. 
Further, \textit{Quantum Fuzzy Neural Networks (QFuzzyNNs)}~\cite{QFL-QFuzzyNN}, inspired by tensor network methodologies, present a unique combination of quantum computing principles and fuzzy logic. 
By utilizing quantum superposition and entanglement, QFuzzyNNs conduct fuzzy inference, enhancing decision-making capabilities even with incomplete information~\cite{8651334}. In Fig.~\ref{fig:Quantum-Counterparts}, we present the advantages and disadvantages of various neural networks when combined with quantum capabilities.  

On a slightly different track, QAOA~\cite{QAOA} has simultaneously established itself as a potent method to address combinatorial optimization problems. QAOA functions by alternating between quantum and classical optimization stages, enabling it to effectively search through vast solution spaces typically beyond the reach of classical algorithms. The principal innovation of QAOA is its application of quantum circuits for state evolution, followed by classical updates to refine the variational parameters~\cite{QFL-QAOA}. This hybrid method has demonstrated remarkable potential in solving the MaxCut problem, consistently surpassing classical algorithms in both convergence speed and solution quality. The capacity of QAOA to provide near-optimal solutions for NP-hard problems underscores its importance in diverse optimization tasks, ranging from logistics~\cite{azzaoui2021quantum,dalal2024digitized} to cryptography~\cite{phab2022first,kashapogu2024exploring}.
Recently, researchers have explored the application of quantum tensors for cyber-physical-social systems~\cite{QFL-Tensor, QFL-Tensor2} and revealed that tensor decomposition techniques can substantially decrease communication overhead in distributed learning settings. 

An inspection of $\mathrm{QUE}$ for quantum-inspired classical algorithms reveals varied occupancy levels tied to circuit depth and qubit usage. FedQNN’s generalized framework (\(\mathrm{Q}_{alloc}=5\), \(L=3\)) attains \(\mathrm{QUE}\approx1.0\), making it well suited to asynchronous updates that exploit its high utilization and low-latency gradient pushes \cite{QFL2}. The QCNN-based privacy model (\(\mathrm{Q}_{alloc}=4\), \(L=2\)) yields \(\mathrm{QUE}=1.0\) and favors synchronous aggregation to ensure consistent feature extraction across non-IID medical nodes \cite{QFL-QCNN}. QLSTM’s deeper memory-enhancing circuits (\(\mathrm{Q}_{alloc}=4\), \(L=4\)) report \(\mathrm{QUE}\approx1.0\) and employ specialized asynchronous updates to balance sequence processing with qubit coherence constraints \cite{QFL-QLSTM}. Despite these strengths, tensor-network hybrids (\(\mathrm{Q}_{alloc}=4\), \(L=5\)) show slightly reduced QUE (\(\approx0.80\)) and require hybrid classical–quantum fusion to reconcile their expressive power with aggregation consistency \cite{QFL-Tensor2}. Limits unique to these hybrid frameworks include the overhead of repeated state re-preparation for classical data encoding.

\noindent \textbf{Key insights (FL on NISQ Devices):}
Our critical analysis offers the following key insights.
\begin{itemize}
\item The integration of advanced optimization techniques for quantum-classical systems is critical to overcoming the limitations of NISQ devices. By intelligently distributing computational tasks, employing ML to predict optimal parameters, and adapting to noise conditions, the researchers can significantly enhance the practicality and scalability of QFL systems.
\item The current hybrid gradient descent techniques lack a thorough exploration of scalability across quantum hardware and potential latency in distributed training. Thus, adaptive techniques for heterogeneous quantum devices and noise resilience need to be developed to bolster large-scale applications.
    \item The creation of quantum-inspired classical algorithms marks a notable advancement in computational science by integrating the benefits of quantum mechanics with traditional algorithmic frameworks. From the groundbreaking QNNs to recent progress in QAOA, tensor networks, and fuzzy logic, these developments underline the expanding influence of quantum computing on classical systems. 
    
\end{itemize}

\noindent Furthermore, Table~\ref{tab:quantum_federated_analysis} concludes with a comprehensive overview of FL research on NISQ devices, including their strengths, limitations, QUE, and training strategies.

\section{Privacy-Preserving QFL}
\label{sec:Privacy-Preserving QFL}
\begin{table*}[h!]
\centering
\caption{Critical analysis of privacy-preserving QFL.}
\begin{tabular}{|c|p{4cm}|p{4cm}|p{3.5cm}|p{3.5cm}|}
\hline
\textbf{Paper} & \textbf{Strengths} & \textbf{Limitations} & \textbf{QUE [Major reason]} & \textbf{QMTS} \\
\hline
\cite{QFL-SecAggPostQ1} 
  & Robust post-quantum security with efficient 3-round protocol; lower computational load 
  & Limited analysis on dropout handling and large-scale scalability 
  & \begin{tabular}[t]{@{}l@{}}Moderate\\[0.3ex]{[}optimized homomorphic\\encryption{]}\end{tabular} 
  & Synchronous aggregation \\
\hline
\cite{QFL-DPFHE} 
  & Integrates fully homomorphic encryption, achieving 40× training efficiency boost; strong AUC performance 
  & High computational overhead of FHE; deployment challenges in NISQ settings 
  & \begin{tabular}[t]{@{}l@{}}High\\[0.3ex]{[}matrix-vector encoding{]}\end{tabular} 
  & Synchronous aggregation \\
\hline
\cite{QFL-DPNoise} 
  & Introduces controlled quantum noise to enhance adversarial robustness; effective on high-dimensional data 
  & Performance degrades beyond optimal noise levels; lacks extensive real-world scaling tests 
  & \begin{tabular}[t]{@{}l@{}}Moderate\\[0.3ex]{[}noise injection strategy{]}\end{tabular} 
  & Asynchronous aggregation \\
\hline
\cite{QFL-BlindC} 
  & Blind quantum computing ensures strong privacy; robust under gradient attacks 
  & High computational cost; dependency on advanced quantum infrastructure 
  & \begin{tabular}[t]{@{}l@{}}Moderate\\[0.3ex]{[}blind computing methods{]}\end{tabular} 
  & Synchronous aggregation \\
\hline
\cite{QFL-AsynchAgg} 
  & Fully asynchronous updates; strong privacy via Paillier encryption 
  & High encryption overhead; limited dropout handling 
  & \begin{tabular}[t]{@{}l@{}}Moderate\\[0.3ex]{[}Paillier-based aggregation{]}\end{tabular} 
  & Asynchronous aggregation \\
\hline
\cite{QFL-DP} 
  & Achieves \(>98\%\) test accuracy with \(\epsilon<1.3\); strong privacy protection 
  & Noise trade-offs impact performance; scaling challenges 
  & \begin{tabular}[t]{@{}l@{}}High\\[0.3ex]{[}noise-resilient VQC{]}\end{tabular} 
  & Asynchronous aggregation \\
\hline
\cite{QFL-GradientDP} 
  & Two distinct protocols with low communication costs; secure gradient protection via quantum states 
  & Practical deployment challenges; noise in quantum communication 
  & \begin{tabular}[t]{@{}l@{}}Moderate\\[0.3ex]{[}efficient state encoding{]}\end{tabular} 
  & Synchronous aggregation \\
\hline
\cite{QFL-SecAggPostQ2} 
  & Lattice-based scheme reduces computational overhead by 20\%, double masking enhances privacy 
  & Multi-stage reconstruction complexity; limited scalability for extensive participants 
  & \begin{tabular}[t]{@{}l@{}}Moderate\\[0.3ex]{[}lattice cryptography{]}\end{tabular} 
  & Synchronous aggregation \\
\hline
\cite{QFL10} 
  & GHZ-based aggregation ensures privacy; low computational complexity without encryption overhead 
  & Dependence on quantum infrastructure; scalability to large datasets has not been demonstrated 
  & \begin{tabular}[t]{@{}l@{}}High\\[0.3ex]{[}GHZ state aggregation{]}\end{tabular} 
  & Synchronous aggregation \\
\hline
\end{tabular}
\label{tab:qfl_critical_analysis}
\end{table*}

Privacy-preserving mechanisms in QFL are essential for safeguarding sensitive information while leveraging the computational advantages of quantum systems. QFL introduces advanced techniques such as quantum differential privacy~\cite{hirche2023quantum, zhou2017differential}, secure multi-party computation~\cite{crepeau2002secure,QFL-SecAggPostQ2}, and homomorphic encryption~\cite{broadbent2015quantum,QFL-DPFHE} to mitigate threats like gradient leakage and model inversion. These methodologies not only enhance data security but also address the unique challenges posed by quantum noise, ensuring robust performance in sensitive applications such as healthcare, finance, and autonomous systems.

\noindent $\bullet$ \textbf{Challenges:} Implementing privacy-preserving mechanisms in QFL presents unique challenges that intertwine the complexities of QC with stringent data confidentiality requirements. A significant concern is the potential for quantum-specific attacks, such as quantum gradient inversion, where adversaries exploit quantum information to reconstruct sensitive data from shared gradients~\cite{larson2020quantum}. Unlike classical systems, the quantum nature of data and computations necessitates novel privacy-preserving protocols that can effectively conceal gradient information without compromising the advantages of quantum processing. 

Moreover, the integration of Differential Privacy (DP) into QFL frameworks is not straightforward. The inherent noise in NISQ devices complicates the application of traditional DP techniques~\cite{zhao2024bridging}. Additionally, the development of quantum-compatible encryption methods, such as quantum homomorphic encryption, remains in its infancy, presenting challenges in performing computations on encrypted data without exposing sensitive information~\cite{fisher2014quantum}. These limitations underscore the need for advanced research into quantum-specific privacy-preserving techniques that can balance the trade-offs between data security and computational efficiency in FL environments.

\subsection{Quantum Differential Privacy in Federated Systems}

Quantum Differential Privacy (QDP) plays a vital role in protecting sensitive data in FL systems, particularly when integrated with quantum computing benefits. A protocol using blind quantum computing has been introduced by~\cite{QFL-BlindC} for private QFL, which utilizes quantum servers for intensive computations and ensures data privacy through differential privacy. However, a drawback is the possible decline in performance when extending this protocol to more intricate datasets and non-IID data distributions.
The work in~\cite{QFL-DP} discussed the challenges of integrating QFL with QDP on NISQ devices and emphasized the importance of balancing privacy preservation with model performance.
An innovative quantum fuzzy FL framework is introduced by~\cite{QFL-QFuzzyNN}, which amalgamates fuzzy logic with quantum FL to augment both privacy and model efficacy. 

Using FHE, the authors in~\cite{QFL-DPFHE} proposed high-performance and quantum-safe federated learning, a vertical FL strategy, to achieve quantum-resistant training. 
One of the most compelling aspects of the study is the substantial boost in training efficiency, reported as up to 40 times faster than conventional methods while maintaining an Area Under the Curve (AUC) comparable to standard algorithms like logistic regression and XGBoost. 
Furthermore, the work~\cite{QFL-DPNoise} investigated the use of quantum noise to improve model robustness in FL, particularly against adversarial attacks.

An assessment of $\mathrm{QUE}$ for privacy‐preserving QFL protocols indicates that frameworks prioritizing encryption‐heavy workflows tend to operate with moderate qubit–layer occupancy (\(\mathrm{QUE}\approx0.6\)–0.8), while those with lighter embedding circuits achieve higher occupancy. For instance, the FHE‐driven vertical update in \cite{QFL-DPFHE} typically runs on a four‐qubit, three‐layer VQC (\(\mathrm{Q}_{alloc}=4\), \(L=3\)), yielding \(\mathrm{QUE}\approx0.67\) and justifying its synchronous aggregation to amortize the cost of homomorphic operations. The differential‐privacy–enhanced protocol in \cite{QFL-DP} uses a similar shallow ansatz (\(\mathrm{Q}_{alloc}=5\), \(L=3\)), achieving \(\mathrm{QUE}\approx0.67\) and favoring asynchronous updates to hide noise patterns across rounds. Blind computing methods in \cite{QFL-BlindC} deploy deeper circuits (\(\mathrm{Q}_{alloc}=6\), \(L=4\)), giving \(\mathrm{QUE}\approx0.75\) and employing synchronous UBQC updates to maintain coherent secret sharing. Noise‐adaptive training in \cite{QFL-DPNoise} uses a three‐qubit, two‐layer ansatz (\(\mathrm{Q}_{alloc}=3\), \(L=2\)), yielding \(\mathrm{QUE}\approx1.0\) and leveraging asynchronous updates to inject and calibrate noise on the fly. Finally, the Paillier‐secured asynchronous framework in \cite{QFL-AsynchAgg} runs on a minimal two‐qubit, two‐layer circuit (\(\mathrm{QUE}=1.0\)), matching its lock‐free aggregation for low‐latency privacy protection. Despite their strengths, these protocols often lack standardized metrics for end-to-end latency and energy cost critical metrics for real-world deployment.

\subsection{Quantum Model Aggregation and Attacks}
The importance of privacy-preserving model aggregation has become pronounced with the continuous expansion of FL, particularly within quantum computing contexts. In~\cite{QFL-AsynchAgg}, a concept of privacy-preserving asynchronous FL has been introduced to facilitate secure model aggregation without necessitating the synchronization of client updates. The authors employed Paillier encryption~\cite{paillier2005paillier} to support the homomorphic aggregation of encrypted model updates. Here, clients can update asynchronously by contributing local models, with the server aggregating these updates without disclosing private information. 
Addressing the need for post-quantum security, the work~\cite{QFL-SecAggPostQ1} presented a secure three-round aggregation protocol to ensure quantum resistance, using post-quantum cryptographic techniques. 

\begin{figure}[h]
    \centering
    \includegraphics[width=1\linewidth]{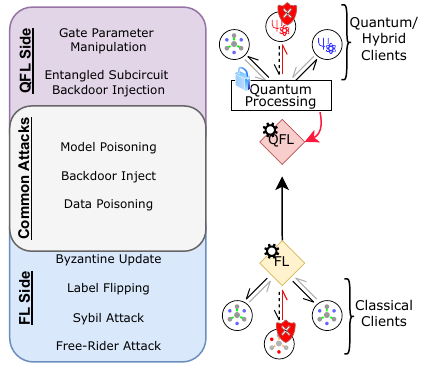}
    \caption{Illustrating attacks on FL and QFL frameworks.}
    \label{fig:enter-label}
\end{figure}
One prominent direction concerns safeguarding FL infrastructures against quantum attacks. Fig.~\ref{fig:enter-label} illustrates various attacks on quantum, FL, and QFL sides. The work~\cite{QFL-SecAggPostQ2} introduced an FL framework that utilizes lattice-based multi-stage secret sharing to ensure resistance against quantum attacks. This scheme employs dual masking techniques to safeguard model parameters, enabling reconstruction of masked model updates while minimizing frequent communication overhead. 
On a similar track, the authors in~\cite{gharavi2025pqbflpostquantumblockchainbasedprotocol} introduced a framework that leverages post-quantum cryptographic primitives to mitigate eavesdropping and model-tampering threats. 
Parallel to security concerns, researchers have explored adversarial defenses within QFL~\cite{maouaki2025qfalquantumfederatedadversarial}, which integrates adversarial training in a quantum federated environment, generating local perturbations at each node and sharing robust gradients via FedAvg. 

From an attack perspective, to counter model poisoning and gate‐parameter manipulation in QFL, operators can adopt Byzantine‐robust aggregation rules and runtime validation of quantum updates. Classical FL defenses such as Krum or coordinate‐wise median have proven effective against malicious client updates \cite{Blanchard2017Krum}, and can be extended to QFL by treating each client’s variational parameter shift as a vector for robust aggregation. Additionally, clients and the server can interleave randomized “test circuits,” which would essentially be short sequences with known output distributions, into the training pipeline. By measuring the fidelity of these test circuits against expected outcomes using the average‐gate‐fidelity formula \cite{QC9}, the server can detect aberrant gate manipulations or backdoor injections before integrating a client’s update.

Further, from an aspect of a distributed denial-of-service (DDoS), the attacker can target three pressure points: the central aggregator, Quantum-Compute-as-a-Service (QCaaS) endpoints for PQC evaluations, and intermediate relays that coordinate submissions and returns. Congestion at any of these points reduces effective throughput, inflates staleness \(\Delta t_k\), and, most critically for privacy, shrinks the instantaneous participant set \(\mathcal{S}_t\). Secure aggregation delivers its guarantees only when the active set meets a minimum threshold; letting rounds proceed with too few contributors erodes anonymity and exposes client updates.

To make the dependency explicit, let \(k_{\min}\) denote the minimum cohort size required by the secure aggregation protocol. A round is privacy-preserving only if \(|\mathcal{S}_t|\ge k_{\min}\); otherwise it must be deferred rather than “fail-open”:
\begin{equation}
\mathrm{proceed}(t) =
\begin{cases}
1, & \text{if } |\mathcal{S}_t| \ge k_{\min},\\
0, & \text{otherwise,}
\end{cases}
\label{eq:hold-until-k}
\end{equation}
This rule decouples privacy loss from traffic volatility. When DDoS induces long queues at QCaaS and raises \(\Delta t_k\), aggregation can remain correct by applying the staleness decay \(w_k(\Delta t_k)\) while still enforcing \eqref{eq:hold-until-k}. The practical failure mode to avoid is a forced reduction of \(k_{\min}\) or a switch to plaintext fallback under load; both actions convert an availability incident into a privacy incident.
Mitigating this involves mainly three layers: 
\begin{enumerate}
    \item \emph{Capacity and scheduling:} Implement redundant aggregators with any-cast, regulate client access via rate limits and quotas, randomize PQC job timing to prevent side-channels, and prioritize ongoing rounds to preserve \(|\mathcal{S}_t|\).
    \item \emph{Transport and anonymity:} Treat relays as honest-but-curious, use anonymous publication and PIR-style access \cite{khan2025controlledpub,khan2022qapir}, manage high-contention with quantum collision detection to thwart adversarial actions \cite{khan2021collision}, and delegate quantum tasks with concealed computation to protect compute nodes' knowledge \cite{zaman2023concealed}. 
    \item  \emph{Content reduction and privacy budgets:} Limit symbol exposure by using meaning-level representations \cite{khalid2023qsemantic}, hold rounds until \(|\mathcal{S}_t|\ge k_{\min}\) and apply secure aggregation, and ensure per-round differential privacy to stay within the budget during attacks.
\end{enumerate}
 For NISQ deployments, addressing queue sensitivity and scarcity, and integrating these controls with the quantum stack from scratch would be an ideal approach \cite{khalid2024quantumnetwork}. DDoS mainly affects QFL availability but also risks privacy via amplified metadata and timing traces.
 
 Beyond focused attacks such as DDoS, QFL deployments are also exposed to several other distributed threats that can erode confidentiality, integrity, and trust such as :
 \begin{itemize}
     \item \textbf{Collusion attacks}: Groups of malicious or semi-honest clients can collaborate to reconstruct global model updates, infer private data, or bias aggregation. In classical FL, this already affects differential privacy; in QFL, the risk is higher due to fewer active quantum clients. Coordinated, correlated PQC updates can also deceive anomaly detectors. Mitigation needs secure aggregation protocols that handle collusion and cryptographically shuffle client identities during aggregation~\cite{Bonawitz2017SecAgg,khan2025controlledpub}.
     
     \item \textbf{Gradient inference and reconstruction}: A curious server or relay may reconstruct training data from gradients. Although quantum noise adds variability, hybrid gradients with classical parts can still leak. Mitigating this involves adding differential privacy noise to both quantum and classical gradients before transmission~\cite{mcmahan2017communication}, along with secure aggregation to obscure individual contributions~\cite{Bonawitz2017SecAgg}.

     \item \textbf{Compromise of relay or scheduler nodes}: Intermediate nodes managing PQC jobs can tamper with updates, leak metadata, or reorder traffic. Since QFL uses third-party QCaaS or shared infrastructure, these nodes are vulnerable to attacks. Using concealed quantum telecomputation~\cite{zaman2023concealed} and anonymous publication mechanisms~\cite{khan2025controlledpub,khan2022qapir} can protect client identity and metadata, preventing attribution or targeted delays.

     \item \textbf{Model drift through stale or low-quality updates}: In asynchronous QFL, stragglers may submit stale updates using outdated global parameters. If unweighted, these updates can detract the global model from the optimum, reducing accuracy or enabling stealth poisoning attacks. This is worsened on NISQ devices due to calibration drift. Techniques like staleness-weighted aggregation~\cite{Xie2019FedAsync} and validation-based reputation scoring can mitigate harmful updates without excluding honest but slow clients.

 \end{itemize}

\noindent $\bullet$ \textbf{Resilience against model and data poisoning:} 
A central question is whether the integration of quantum resources alters the resilience of QFL against widely-known threats such as model and data poisoning. In the classical setting, model poisoning attacks involve adversarial clients submitting deliberately corrupted gradients or updates with the goal of steering global convergence toward an incorrect or backdoored solution. In QFL, a similar threat manifests through the manipulation of variational parameters within PQCs or the injection of malicious amplitude encodings. Although the no-cloning theorem prevents adversaries from perfectly replicating quantum states to amplify their influence, the aggregation rule remains vulnerable to skewed updates~\cite{9951322}. Thus, QFL does not eliminate the fundamental risk of model poisoning, but rather shifts its locus from tensor weights to quantum parameters. 

Overall, QFL inherits the poisoning vulnerabilities of classical FL while introducing additional complications due to limited error-correction capacity and the opacity of quantum encodings. Yet it also offers distinctive points of resilience: (i) the impossibility of exact state replication restricts large-scale poisoning amplification, and (ii) cryptographic overlays combined with quantum verification techniques can enforce stronger guarantees of update integrity. Consequently, the resilience of QFL against poisoning is not absolute but conditional, hinging on careful protocol design and the alignment of aggregation rules with the constraints of NISQ devices. A comparative overview of vulnerabilities in classical FL versus QFL is summarized in Table~\ref{tab:qfl_vs_fl_risks}.

\begin{table*}[h]
\centering
\setlength{\tabcolsep}{5pt}
\caption{Comparison of vulnerabilities of FL and QFL against various attacks. QFL tends to be more ($\uparrow$) or similar ($\approx$) vulnerable than classical FL.}
\begin{tabular}{p{3cm} c p{6.5cm} p{5.8cm}}
\toprule
\textbf{Risk class} & \textbf{QFL vs FL} & \textbf{In QFL: Implementation; Challenges} & \textbf{Mitigation Strategy for QFL} \\
\midrule

\textbf{Model poisoning} & $\uparrow$ &
Manipulation of PQC parameters and encoder calibrations can bias global updates; NISQ noise may mask anomalous drifts, complicating detection~\cite{khalid2024quantumnetwork}. &
Robust/Byzantine aggregation; PQC-parameter anomaly detection; hybrid-key secure aggregation \cite{Blanchard2017Krum,Yin2018Byz,Bonawitz2017SecAgg} \\
\addlinespace[2pt]
\textbf{Backdoor inject} & $\approx$ &
Gate-parameter steering and trigger states embedded in variational circuits; deeper NISQ circuits increase susceptibility without full QEC~\cite{khalid2024quantumnetwork}. &
Trigger sanitization; parameter attestation for PQC layers; cross-round consistency checks. \cite{zaman2023concealed} \\
\addlinespace[2pt]
\textbf{Data poisoning} & $\approx$ &
Preparation of corrupted quantum states or adversarial superpositions; heterogeneous noise obscures intent, degrading screening~\cite{khalid2024quantumnetwork}. &
Outlier clipping; robust aggregation; input-state audits / spot verification of measurement statistics. \cite{ Yin2018Byz,Blanchard2017Krum, khalid2024quantumnetwork}\\
\addlinespace[2pt]
\textbf{Eavesdropping} & $\uparrow$ &
Extra hops to QCaaS and relays expose timing, routing, and job metadata; intermediate nodes may be honest-but-curious~\cite{zaman2023concealed,khan2025controlledpub,khan2022qapir,khan2021collision}. &
Concealed telecomputation; anonymous publication / PIR; authenticated channels; relay anonymity. \cite{khan2025controlledpub,khan2021collision} \\
\addlinespace[2pt]
\textbf{Availability (DDoS)} & $\uparrow$ &
Scarce QPU capacity and long queues make training sensitive to traffic floods; reduced active cohort threatens secure-aggregation thresholds~\cite{khalid2024quantumnetwork}. &
Redundant aggregators; admission control and rate limits; hold-until-$k_{\min}$ policy; staleness-aware async updates. \cite{khalid2024quantumnetwork, Bonawitz2017SecAgg}  \\

\bottomrule
\end{tabular}
\label{tab:qfl_vs_fl_risks}
\end{table*}

Quantum‐specific error mitigation techniques further strengthen security by masking subtle parameter perturbations. Techniques like zero‐noise extrapolation or probabilistic error cancellation \cite{Cerezo2021} can be applied selectively to aggregated model parameters, reducing the impact of malicious noise injections while preserving the integrity of genuine updates. From a cryptographic standpoint, integrating verifiable blind QC protocols \cite{QFL-BlindC} allows the server to challenge clients with hidden circuit specifications, guaranteeing that parameterized unitaries are executed as intended without revealing the model’s secrets.  Together, these measures: robust aggregation, fidelity‐based auditing, error‐mitigation filtering, and verifiable execution, offer a layered defense tailored to the hybrid vulnerabilities of QFL.  

An assessment of \(\mathrm{QUE}\) for secure‐aggregation and attack‐resilient QFL schemes shows uniformly high occupancy due to their use of shallow embedding circuits. For example, the Paillier‐based asynchronous framework in \cite{QFL-AsynchAgg} runs on a minimal two‐qubit, two‐layer ansatz (\(\mathrm{Q}_{alloc}=2\), \(L=2\)), yielding \(\mathrm{QUE}=1.0\) and matching its lock‐free aggregation strategy. Post‐quantum MPC in \cite{QFL-SecAggPostQ1} and lattice‐masked secret sharing in \cite{QFL-SecAggPostQ2} both employ four‐qubit, three‐layer circuits (\(\mathrm{Q}_{alloc}=4\), \(L=3\)) with \(\mathrm{QUE}=1.0\), justifying synchronous updates to amortize cryptographic overhead. The blockchain‐anchored protocol in \cite{gharavi2025pqbflpostquantumblockchainbasedprotocol} uses a three‐qubit, two‐layer design (\(\mathrm{QUE}=1.0\)) and also opts for synchronous aggregation to ensure ledger consistency. The adversarial defense in \cite{maouaki2025qfalquantumfederatedadversarial} leverages a four‐qubit, four‐layer QNN (\(\mathrm{QUE}=1.0\)) and employs asynchronous FedAvg with local perturbation to balance robustness and throughput. Despite their strong QUE, these protocols share key limitations. These models generally ignore client dropout and dynamic participation, lack adaptive privacy‐budget tuning, and incur substantial computational and communication costs from heavy cryptographic primitives. Future work should explore lightweight quantum‐safe primitives, robust dropout‐tolerant aggregation, and real‐time privacy budget adjustment to enhance scalability and resilience in practical QFL deployments.

\subsection{Quantum Secure Multi-Party Computation}

In conventional FL systems, challenges such as gradient inversion attacks, which enable the reconstruction of private gradients, are prevalent. 
The authors in~\cite{QFL-GradientDP} addressed this issue by utilizing quantum states to encode gradients, thereby securing multi-party contributions to the global model from adversaries. The proposed framework incorporates protocols that use private inner-product estimation and incremental learning, collectively allowing contributions from each party without exposing individual gradients. In another work~\cite{QFL10}, a quantum secure aggregation scheme is proposed to facilitate secure multi-party contributions by aggregating local models using quantum secret sharing, where participants encode model gradients into qubits and transmit them via Greenberger–Horne–Zeilinger (GHZ) states~\cite{greenberger2009ghz}. Quantum entanglement detects eavesdropping, ensuring data confidentiality. 
Further, in~\cite{QFL-HQK}, a hybrid-quantum-key based secure FL system is designed for multi-party applications, particularly in the medical field. This system uses quantum key distribution to securely share parameters and hybrid quantum keys to protect data exchanges and model contributions from external attacks. 

\noindent $\bullet$ \textbf{Untrusted but non-adversarial computation nodes:}
Many QFL deployments involve service providers or benign participants that execute parts of the workflow but cannot be granted access to plaintext updates or model states, such as an outsourced PQC execution on a cloud QPU provider or aggregation nodes with stricter data-handling policies. These nodes need not necessarily be an attacker and thus must be carefully considered in the QFL framework. 

A practical approach is to delegate quantum subroutines while concealing inputs and program structure from the compute node. It can be done by arranging the interaction so that an untrusted server performs the required quantum operations yet learns negligible information about the task beyond coarse resource usage. Such protocols have been proposed for low-latency, reliability-critical settings and are directly compatible with PQC calls within a QFL round~\cite{zaman2023concealed}. When QFL relies on intermediate routing or publication layers, anonymous transport primitives help prevent metadata leakage even if those layers are honest-but-curious. Recent results on controlled quantum anonymity ~\cite{khan2025controlledpub}, quantum anonymous private information retrieval in distributed networks~\cite{khan2022qapir}, and quantum anonymous collision detection for coordinating access~\cite{khan2021collision} provide building blocks for hiding the association between a client and its update/query while maintaining accountability and liveness in multi-party environments. In bandwidth-limited regimes, quantum semantic communication has also been explored to reduce exposure of raw symbols by transmitting meaning-level representations over shared links, which can lower leakage through intermediate nodes when combined with keying and aggregation~\cite{khalid2023qsemantic}.

For QFL, these primitives map cleanly onto common roles. If the compute node is untrusted, clients keep data and classical parameters locally, delegate only the PQC evaluations with concealed inputs, and request verifiable transcripts of the delegated rounds for audit. If the aggregator is untrusted, clients retain their secure aggregation scheme and hybrid-keying and additionally route updates anonymously to hide which client contributed a given update. If aggregation nodes are untrusted, anonymous collision-detection and semantic communication reduce side-channel leakage while preserving throughput.

\noindent $\bullet$ {\bf Anonymous-network extensions:} A distinct yet emerging strand of work complements quantum secure multi-party computation by focusing on anonymity within quantum networks. Mechanisms such as controlled quantum anonymous publication~\cite{khan2025controlledpub}, quantum anonymous private information retrieval for distributed networks~\cite{khan2022qapir}, and quantum anonymous collision detection for coordinating access to shared channels~\cite{khan2021collision} provide building blocks for concealing client identities and query patterns while maintaining verifiability. In parallel, concealed quantum telecomputation has been proposed for ultra-reliable low-latency communication contexts, enabling delegated computation on untrusted infrastructure without revealing inputs or program structure~\cite{zaman2023concealed}. Incorporating such primitives into QFL frameworks would extend beyond confidentiality of data values to protection of user metadata and communication traces, thereby strengthening privacy guarantees under both honest-but-curious and active adversarial models. This trajectory also opens avenues for future research on privacy-centric designs, where federated optimization is augmented with anonymity-preserving layers to reduce information leakage through routing and scheduling.

Finally, Analyzing \(\mathrm{QUE}\) for secure multi‐party computation schemes indicates near‐maximal qubit utilization across protocols. The inner‐product encoding in \cite{QFL-GradientDP} uses a four‐qubit, three‐layer circuit (\(\mathrm{Q}_{alloc}=3\), \(L=3\)) obtaining \(\mathrm{QUE}=1.0\) and thus adopts synchronous aggregation to leverage full‐batch quantum metric estimation; the GHZ‐based secret sharing in \cite{QFL10} runs on a three‐qubit, two‐layer design (\(\mathrm{Q}_{alloc}=3\), \(L=2\)) with \(\mathrm{QUE}=1.0\) and similarly favors synchronous updates to maintain entanglement integrity; and the quantum‐key DFL in \cite{QFL-HQK} employs a five‐qubit, four‐layer ansatz (\(\mathrm{Q}_{alloc}=5\), \(L=4\)) to achieve a high \(\mathrm{QUE}\approx1.0\) paired with hybrid classical–quantum fusion to balance key‐distribution overhead with model convergence. Despite their efficient parameter‐gate ratios and aligned model strategy choices, these multi-party computation frameworks have significant limitations. Firstly, they assume ideal entanglement distribution and noise‐free communication, omit analysis under non‐IID data splits, and lack adaptive mechanisms for client dropout or key‐renewal scheduling. Future work should explore decoherence‐aware aggregation protocols, resilient key management strategies in dynamic networks, and benchmark suites that jointly measure privacy, QUE, and end‐to‐end latency in dynamic QFL environments. \\

\noindent \textbf{Key insights (Privacy-Preserving QFL):} With the critical review of the existing work on Privacy-preserving QFL, we have come across the following key insights.
\begin{itemize}
\item  
Although noteworthy progress has been achieved in merging QDP with quantum FL, important obstacles persist. Particularly, the balance between privacy and accuracy, the scalability of quantum systems, and the effects of noise in practical applications are topics necessitating further research. Overcoming these challenges is vital for the widespread implementation of QDP in federated systems.
\item Despite good progress in enhancing privacy-preserving quantum model aggregation, several critical challenges are yet to be explored, such as quantum noise, computational overhead, and the trade-offs between privacy and performance. It is imperative to address these issues to ensure the scalability and robustness of these solutions within FL environments. Subsequent research should prioritize the optimization of QFL protocols for heterogeneous and large-scale systems, with a focus on mitigating the constraints imposed by current quantum hardware.
    
\item   
The literature demonstrates how quantum techniques such as gradient hiding, GHZ state entanglement, and quantum key distribution can be integrated into FL to protect the contributions of all participants. While these methods show promising results, they must address challenges related to quantum noise and resource constraints to be viable in real-world multi-party FL systems.
\end{itemize}

\noindent Finally, Table~\ref{tab:qfl_critical_analysis} offers a comprehensive analysis of privacy-preserving QFL research, detailing the strengths, limitations, QUE, and training strategies employed in various works.

\section{Miscellaneous}
\label{sec:Miscellaneous}
This section aims to elaborate on the practical usage and implementation of QFL, while discussing studies that apply it to real-life datasets that demonstrate QFL's true potential. It also discusses the trainability and learnability aspects of QFL to further discuss challenges and opportunities. 
\subsection{Real-life Applications of QFL}
\begin{figure}[h]
    \centering
    \includegraphics[width=1\linewidth]{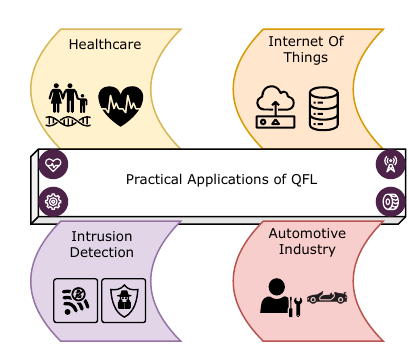}
    \caption{Broad application domains of QFL research. }
    \label{fig:misc}
\end{figure}

Numerous studies have also explored QFL to enhance performance and address the computational demands in diverse application domains. Fig.~\ref{fig:misc} shows application domains where QFL research has shown its footprints.

In {\bf healthcare contexts}, federated quantum designs often emphasize data confidentiality. One dementia-classification method \cite{tanbhir2025quantuminspiredprivacypreservingfederatedlearning} employs QKD for encrypting parameter updates and safeguarding sensitive MRI data across hospital nodes. Despite a 1\% increment in computational overhead, classification accuracy matched the baseline CNN (roughly 94\%) while guaranteeing that eavesdroppers cannot easily invert gradients. Another proposal targets secure intrusion detection in consumer networks \cite{10679217}, achieving up to 3--4\% gains in detection rates on complex traffic features when harnessing small-scale QNNs and homomorphic gradient aggregations.
Addressing data heterogeneity is a central challenge in large-scale QFL. The Dynamic Aggregation QFL (DAQFL) approach \cite{DAQFL} introduces an adaptive weighting scheme based on local accuracy improvements. 
In healthcare IoT benchmarks, DAQFL achieves a 2--6\% accuracy gain over baseline quantum federated averaging when dealing with skewed or long-tailed data distributions. QFL has also shown efficacy for sensitive clinical tasks such as pain-level detection. In~\cite{balasubramani2025novel}, ECG signals are processed with a quantum-transfer-learning pipeline, integrated within a federated scheme to preserve patient privacy. Compared to conventional CNN-based methods, the quantum model improved accuracy by nearly 2\%, reaching approximately 94.8\% for multi-tier pain labeling. 

Moving into {\bf automotive applications}, QFSM \cite{qu2024qfsm} addresses speech-emotion recognition in vehicular networks by employing quantum minimal-gated recurrent units. This design shortens the gating operations and cuts training convergence time by roughly 10--15\%. Field tests in a 5G IoV context report up to 2\% higher accuracy than classical gated recurrent units on large-scale SER corpora, while also offering improved noise robustness in fluctuating channel conditions.
Vehicular networks have also been extended into metaverse-based coordination. A quantum-enhanced FL framework \cite{10758814} uses hierarchical clustering of vehicular nodes, where QNN states are compressed via quantum principal-component analysis. Adaptive reinforcement scheduling then modulates local updates, ensuring resource balancing across immersive metaverse tasks. Simulation results reflect around 15\% lower communication overhead compared to standard FL.
A related thread investigates quantum FL in edge or drone-assisted scenarios. One surveillance solution \cite{10829961} allows UAVs to transmit partial circuit parameters, dynamically adjusting QNN layer depths based on instantaneous channel quality. This reduces the communication load by nearly 20\% while retaining top-1 accuracy comparable to a static-depth model on an aerial license-plate dataset. 

Another notable approach, ESQFL \cite{10734228}, integrates digital twin frameworks for real-time {\bf power-grid stability} analysis. By linking quantum FL updates to digital replicas of physical bus nodes, the global model can detect incipient voltage collapse and dynamically schedule corrective actions. Shapley value calculations explain each node’s contribution in an interpretable manner. Grid simulations showed a 4\% improvement in detecting instability events and a 10\% boost in overall reliability.

Beyond the model-training process, some works explore hybrid optimization for FL resource management. One Benders’ decomposition algorithm~\cite{10636852} splits scheduling of workers and continuous learning-rate decisions into subproblems, solved alternately on a quantum annealer and a classical solver. At moderate scales, the method reduces runtime by up to 70\% versus purely classical MILP solutions, indicating promise for large-scale FL scheduling.
A further refinement involves quantum gate customization. Entanglement-controlled QFL \cite{10878965} employs specialized universal gates to fine-tune the entanglement entropy in multi-depth QNNs. This strategy mitigates the barren-plateau problem by balancing inter-qubit interference. Experiments indicate consistent convergence on non-iid data with minimal gradient vanishing.

Altogether, these approaches underscore a rapidly maturing QFL ecosystem, increasingly oriented toward privacy-preserving, high-accuracy, and resource-efficient learning in complex distributed environments.

\subsection{Trainability and Learnability of QFL}
\label{sec:Trainability and Learnability of QFL}

Federated quantum models combine the expressive power of parameterized quantum circuits with the decentralized optimization of FL. Yet they inherit two intertwined theoretical challenges: ensuring effective optimization (\emph{\textbf{trainability}}) and guaranteeing robust generalization (\emph{\textbf{learnability}}).

As circuit depth \(L\) or qubit count \(Q\) increases, gradient magnitudes can vanish exponentially, a phenomenon known as the \emph{barren plateau} \cite{mcclean2018barren,Cerezo2021}. In QFL, each client’s local update may fall below machine noise, forcing the server to perform many more communication rounds to recover signal \cite{QFL-QNGD}. Empirical results on hardware-efficient ansatz show that limiting depth to \(L\le4\) sustains gradients above numeric precision for up to 16 qubits, reducing required rounds by 30–50\% compared to deeper designs \cite{abbas2021power}.
Quantifying convergence, recent studies measure the number of gradient‐descent steps or federated rounds needed to reach a target accuracy. For example, FQNGD achieves 90\% of its asymptotic accuracy within 20 rounds on standard benchmarks, compared to 40 rounds for FedAvg with the same ansatz \cite{QFL-QNGD}. This halving of communication cost stems from \emph{quantum natural gradient} descent, which preconditions updates by the inverse quantum Fisher information matrix, enhancing per‐round progress up to 2× on ideal simulators \cite{QFL-QNGD}. 
Additional acceleration techniques further improve trainability.  \emph{Layerwise training} optimizes one ansatz block at a time and preserves the gradient signal by avoiding global barren plateaus. Alternatively, \emph{smart initialization} reuses parameters from smaller or related circuits to jump‐start optimization, cutting VQA iterations by up to 60\% in QAOA settings \cite{ML-Optimization,skolik2023quantum}. Together, these methods yield hybrid protocols that converge in fewer than 25 rounds even under moderate noise, making QFL practical on NISQ hardware. 
 
On the other hand, learnability examines how well a QFL model trained on local data samples extends to unseen inputs. Classical FL theory derives generalization bounds via VC‐dimension and algorithmic stability; quantum analogues are emerging. Shallow, low‐entanglement circuits behave like kernel machines with controlled Rademacher complexity, suggesting that \(\mathcal{O}(\log N)\) qubits suffice to generalize on \(N\) training examples \cite{schuld2021effect}. Yet over‐parameterized quantum models exhibit a \emph{double descent} curve i.e., after an interpolation threshold, additional parameters can improve generalization under appropriate regularization \cite{Caro_2022, PRXQuantum.2.040321}.

Bounding the number of quantum parameters or qubits for a desired error remains an open problem.  Preliminary results indicate that federated Rademacher complexity of hybrid models scales with the sum of classical and quantum parameter norms, and can be bounded by Eq. \ref{eqn:rademacher}.

\begin{equation}  
\mathcal{R}(\mathcal{F}) \;\le\; \frac{C_{\rm classical}}{\sqrt{n_{\rm classical}}} + \frac{C_{\rm quantum}}{\sqrt{n_{\rm quantum}}},\!
\label{eqn:rademacher}
\end{equation}
\noindent where \(n_{\rm classical}\), \(n_{\rm quantum}\) are local sample sizes and \(C_{\rm classical},C_{\rm quantum}\) depend on circuit expressivity \cite{bu2021rademacher}.  Stability‐based bounds further relate the sensitivity of the global model to perturbations in client updates, with quantum algorithmic stability results implying that small‐norm updates preserve generalization under noisy mixing \cite{yang2025stabilitygeneralizationquantumneural}.

Fully characterizing the \emph{federated barren‐plateau frontier} requires mapping how depth \(L\), width \(Q\), and client heterogeneity jointly determine gradient norms and convergence rates.  On the learnability side, deriving optimal VC‐dimension and Rademacher‐complexity bounds for hybrid quantum–classical hypothesis classes would guide architecture design.
Further work could implement parallelization strategies, efficient qubit routing, or dynamic circuit compilation (based on the device's states) to improve convergence.

\subsection{Hardware Efficiency and Middleware Integration}
\label{sec:Hardware Efficiency and Middleware Integration}

Building on the insights into trainability and learnability, it becomes clear that algorithmic refinements alone cannot unlock the full potential of QFL, without a deep grasp of underlying hardware behavior. NISQ devices impose hard ceilings on qubit count (5-100 qubits), coherence times (10–200$\mu$s), and gate fidelities (90–99\%), all of which vary unpredictably~\cite{Preskill2018,chen2024nisq}. For instance in a vehicular metaverse scenario, quantum-enhanced FL clients running on 10-qubit processors experienced up to 30\% accuracy drift when two-qubit error rates rose above $10^{-3}$, underscoring the sensitivity of federated convergence to hardware noise~\cite{10758814}.

\begin{figure}[h]
  \centering
  \includegraphics[width=\linewidth]{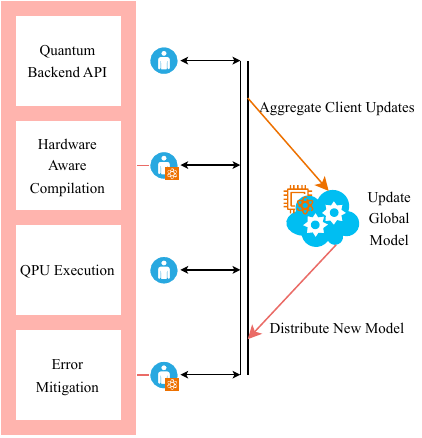}
  \caption{Illustration of middleware pipeline for quantum and hybrid classical-quantum clients.}
  \label{fig:middleware}
\end{figure}

As illustrated in Fig. \ref{fig:middleware}, quantum middleware contains many points of bottlenecks and delays. Queue delays on public QPUs often span 10–60s per job, while compilation, including routing and gate‐reduction passes, can add 100–300ms per circuit \cite{qiskit,10636852}. Heterogeneous client capabilities further distort global updates. Some nodes compile and execute circuits in under a second, others take tens of seconds, leading to stale gradients and model skew \cite{10878965}. Meanwhile, real‐time error rates can drift by 5–10\% over a single training day, mandating on‐the‐fly recalibration and fidelity monitoring \cite{error1}.

Looking forward, co‐design strategies seem promising to reduce these bottlenecks. Dynamic qubit remapping algorithms can reduce SWAP overhead by up to 25\%, preserving circuit depth under tight connectivity constraints \cite{khandavilli2023towards,nash2020quantum}. Embedding error‐mitigation routines such as randomized compiling or zero‐noise extrapolation directly into the compilation pipeline, can recover 50–70\% of lost fidelity without full quantum error correction \cite{Cerezo2021}. From an architectural perspective, a federated hardware abstraction layer may be incorporated that could monitor live calibration metrics and queue statuses, enabling QFL frameworks to schedule gates and adjust ansatz depth automatically based on each client’s current noise profile \cite{ferrari2021compiler}. These directions chart a course toward hardware‐aware QFL that adapts in real time, scales across diverse devices, and sustains federated convergence even as NISQ platforms evolve.

\section{Insights to Experimental Evaluation}
\label{sec:Experimental}

\begin{table*}[h]
\renewcommand{\arraystretch}{1.03}
\caption{Experimental details of recent and benchmark contributions.}
\resizebox{1\textwidth}{!}{
\begin{tabular}{cccccccc}
\hline
\rowcolor[HTML]{F3F3F3} 
\multicolumn{1}{|c|}{\cellcolor[HTML]{F3F3F3}\textbf{Paper}}                      & \multicolumn{1}{c|}{\cellcolor[HTML]{F3F3F3}\textbf{\begin{tabular}[c]{@{}c@{}} non-IID \\ Data \end{tabular}}} & \multicolumn{1}{c|}{\cellcolor[HTML]{F3F3F3}\textbf{Dataset(s)}} & \multicolumn{1}{c|}{\cellcolor[HTML]{F3F3F3}\textbf{\# Clients}} & \multicolumn{1}{c|}{\cellcolor[HTML]{F3F3F3}\textbf{\begin{tabular}[c]{@{}c@{}}\# Training \\ Rounds\end{tabular}}} & \multicolumn{1}{c|}{\cellcolor[HTML]{F3F3F3}\textbf{\begin{tabular}[c]{@{}c@{}}Aggregation\\  Method\end{tabular}}} & \multicolumn{1}{c|}{\cellcolor[HTML]{F3F3F3}\textbf{Accuracy}}                   & \multicolumn{1}{c|}{\cellcolor[HTML]{F3F3F3}\textbf{Libraries}}                                                   \\ \hline
\multicolumn{1}{|c|}{\cite{QFL-TL}}                              & \multicolumn{1}{c|} {\ding{56}}                                           & \multicolumn{1}{c|}{Makecircles\cite{makecircles_sklearn}}                                                                                                                              & \multicolumn{1}{c|}{-}                                           & \multicolumn{1}{c|}{100}                                                                                            & \multicolumn{1}{c|}{\begin{tabular}[c]{@{}c@{}}  - \\ \end{tabular}}                                                                                              & \multicolumn{1}{c|}{91\%}                                                        & \multicolumn{1}{c|}{-}      

\\ \hline

\multicolumn{1}{|c|}{\cite{QFL-TL-2}}                            & \multicolumn{1}{c|}{\ding{56}}                                           & \multicolumn{1}{c|}{\begin{tabular}[c]{@{}c@{}}CIFAR-10\cite{krizhevsky2009learning}, \\ Cats vs dogs\cite{cats_vs_dogs_kaggle}\end{tabular}}                                                                        & \multicolumn{1}{c|}{100}                                         & \multicolumn{1}{c|}{100}                                                                                            & \multicolumn{1}{c|}{FedAvg \cite{mcmahan2017communication}}                                                                                         & \multicolumn{1}{c|}{\begin{tabular}[c]{@{}c@{}} 93\% \\ 98\% \\ \end{tabular}}                                                  & \multicolumn{1}{c|}{\begin{tabular}[c]{@{}c@{}}Qulacs~\cite{qulacs}, \\ PyTorch~\cite{pytorch}, \\ PennyLane~\cite{pennylane}\end{tabular}}                     \\ \hline
\multicolumn{1}{|c|}{\cite{FQE1}}                                & \multicolumn{1}{c|}{\ding{56}}                                           & \multicolumn{1}{c|}{Synthetic}                                                                                                   & \multicolumn{1}{c|}{30}                                          & \multicolumn{1}{c|}{-}                                                                                              & \multicolumn{1}{c|}{FedAvg \cite{mcmahan2017communication}}                                                                                         & \multicolumn{1}{c|}{98\%}                                                        & \multicolumn{1}{c|}{\begin{tabular}[c]{@{}c@{}}Cirq~\cite{cirq}, \\ Tensorflow Quantum~\cite{tfq}, \\ Tensorflow Federated~\cite{tff}\end{tabular}} \\ \hline
\multicolumn{1}{|c|}{\cite{QFL-QNGD}}                            & \multicolumn{1}{c|}{\ding{56}}                                           & \multicolumn{1}{c|}{MNIST\cite{lecun1998gradient}}                                                                                                                                    & \multicolumn{1}{c|}{6}                                           & \multicolumn{1}{c|}{200}                                                                                            & \multicolumn{1}{c|}{FedAvg \cite{mcmahan2017communication}}                                                                                         & \multicolumn{1}{c|}{99\%}                                                        & \multicolumn{1}{c|}{-}                                                                                            \\ \hline
\multicolumn{1}{|c|}{\cite{QFL-QAOA}}                            & \multicolumn{1}{c|}{\ding{52}}                                       & \multicolumn{1}{c|}{MNIST\cite{lecun1998gradient}}                                                                                                                                    & \multicolumn{1}{c|}{-}                                           & \multicolumn{1}{c|}{20}                                                                                             & \multicolumn{1}{c|}{-}                                                                                              & \multicolumn{1}{c|}{81\%}                                                        & \multicolumn{1}{c|}{\begin{tabular}[c]{@{}c@{}}Tensorflow Federated~\cite{tff}, \\ Tensorflow Quantum~\cite{tfq}\end{tabular}}          \\ \hline

\multicolumn{1}{|c|}{\cite{QFL-Tensor2}} & \multicolumn{1}{c|}{\ding{56}}                   & \multicolumn{1}{c|}{\begin{tabular}[c]{@{}c@{}}Synthetic, \\ ECG dataset\cite{mitbih_arrhythmia}, and \\ Diabetes Health Indicator\cite{pima_indians_diabetes_uci}\end{tabular}}      & \multicolumn{1}{c|}{-}                    & \multicolumn{1}{c|}{-}                                                                       & \multicolumn{1}{c|}{Consensus Average \cite{olfati2007consensus}}                                                      & \multicolumn{1}{c|}{-}                                    & \multicolumn{1}{c|}{Tensor Toolbox (Matlab)~\cite{tensorToolbox}}                                              \\ \hline
\multicolumn{1}{|c|}{\cite{QFL-Tensor}}                          & \multicolumn{1}{c|}{\ding{52}}                                       & \multicolumn{1}{c|}{\begin{tabular}[c]{@{}c@{}}RSNA chest X-ray\cite{rsna_chest_xray_pneumonia}, \\ NIH chest X-ray\cite{wang2017chestxray8}, \\ ADNI MRI-scan\cite{adni}, \\ Kidney CT-scan\cite{kits19_data}\end{tabular}} & \multicolumn{1}{c|}{-}                                           & \multicolumn{1}{c|}{35}                                                                                             & \multicolumn{1}{c|}{FedAvg \cite{mcmahan2017communication}}                                                                                         & \multicolumn{1}{c|}{91\%--98\%}                                                  & \multicolumn{1}{c|}{\begin{tabular}[c]{@{}c@{}}Tensorflow Quantum~\cite{tfq}, \\ Tensor Network Library~\cite{tensornetwork}\end{tabular}}        \\ \hline

\multicolumn{1}{|c|}{\cite{QFL-QLSTM}}                           & \multicolumn{1}{c|}{\ding{56}}                                           & \multicolumn{1}{c|}{\begin{tabular}[c]{@{}c@{}}Bessel Function, \\ Struve Function, \\ Delayed Quantum Control\end{tabular}}                                  & \multicolumn{1}{c|}{5}                                           & \multicolumn{1}{c|}{15}                                                                                             & \multicolumn{1}{c|}{FedAvg \cite{mcmahan2017communication}}                                                                                         & \multicolumn{1}{c|}{-}                                                           & \multicolumn{1}{c|}{-}                                                                                            \\ \hline

\multicolumn{1}{|c|}{\cite{QFL-QFuzzyNN}}                        & \multicolumn{1}{c|}{\ding{52}}                                       & \multicolumn{1}{c|}{\begin{tabular}[c]{@{}c@{}}MNIST\cite{lecun1998gradient}, \\ COVID-19\cite{covid_chestxray_dataset} \\ \end{tabular} }                                                                                                                          & \multicolumn{1}{c|}{10}                                          & \multicolumn{1}{c|}{50}                                                                                             & \multicolumn{1}{c|}{\begin{tabular}[c]{@{}c@{}}Quantum \\ Federated \\ Inference \cite{qfi2020}\end{tabular}}                      & \multicolumn{1}{c|}{\begin{tabular}[c]{@{}c@{}} 92\%, \\ 98\% \\ \end{tabular}}                                                  & \multicolumn{1}{c|}{\begin{tabular}[c]{@{}c@{}}PyTorch~\cite{pytorch}, \\ PennyLane~\cite{pennylane}\end{tabular}}                                \\ \hline

\multicolumn{1}{|c|}{\cite{QFL2}}                                & \multicolumn{1}{c|}{\ding{56}}                                           & \multicolumn{1}{c|}{\begin{tabular}[c]{@{}c@{}} Iris\cite{fisher1936iris} ,\\ Breast cancer, DNA\cite{dna_splice_junction_uci} \end{tabular} }                                                                                                                 & \multicolumn{1}{c|}{5}                                           & \multicolumn{1}{c|}{100}                                                                                            & \multicolumn{1}{c|}{\begin{tabular}[c]{@{}c@{}}Riemannian \\ Averaging~\cite{karcher1977riemannian}\end{tabular}}                                & \multicolumn{1}{c|}{\begin{tabular}[c]{@{}c@{}}86\%,\\  90\%\end{tabular}} & \multicolumn{1}{c|}{\begin{tabular}[c]{@{}c@{}} PennyLane~\cite{pennylane},\\  PyTorch~\cite{pytorch} \end{tabular} }                                                                           \\ \hline
\multicolumn{1}{|c|}{\cite{QFL-QCNN}}                            & \multicolumn{1}{c|}{\ding{52}}                                       & \multicolumn{1}{c|}{\begin{tabular}[c]{@{}c@{}} MedNIST\cite{medmnistv2},\\ COVID-19\cite{covid_chestxray_dataset} \\ \end{tabular} }                                                                                                                        & \multicolumn{1}{c|}{10--50}                                      & \multicolumn{1}{c|}{1950}                                                                                           & \multicolumn{1}{c|}{\begin{tabular}[c]{@{}c@{}}FedAvg \cite{mcmahan2017communication}, \\ FedSGD\cite{mcmahan2017communication}\end{tabular}}                                      & \multicolumn{1}{c|}{\begin{tabular}[c]{@{}c@{}} 96\%, \\ 97\% \\ \end{tabular}}                                                  & \multicolumn{1}{c|}{Tensorflow Quantum~\cite{tfq}}                                                                           \\ \hline
\multicolumn{1}{|c|}{\cite{QFL-DPFHE}}                           & \multicolumn{1}{c|}{\ding{56}}                                           & \multicolumn{1}{c|}{Synthetic}                                                                                                                        & \multicolumn{1}{c|}{5}                                           & \multicolumn{1}{c|}{-}                                                                                              & \multicolumn{1}{c|}{HQsFL\cite{QFL-DPFHE}}                                                                                          & \multicolumn{1}{c|}{-}                                                           & \multicolumn{1}{c|}{\begin{tabular}[c]{@{}c@{}}Scikit-Learn~\cite{scikit-learn}, \\ Tensorflow~\cite{tensorflow}, PyTorch~\cite{pytorch}\end{tabular}}                 \\ \hline
\multicolumn{1}{|c|}{\cite{QFL-DPNoise}}                         & \multicolumn{1}{c|}{\ding{56}}                                           & \multicolumn{1}{c|}{\begin{tabular}[c]{@{}c@{}} MNIST\cite{lecun1998gradient}, \\ Moons\cite{make_moons_sklearn}, Synthetic \end{tabular} }                                                                                                                  & \multicolumn{1}{c|}{3}                                           & \multicolumn{1}{c|}{50}                                                                                             & \multicolumn{1}{c|}{-}                                                                                              & \multicolumn{1}{c|}{96\%}                                                        & \multicolumn{1}{c|}{\begin{tabular}[c]{@{}c@{}} PennyLane~\cite{pennylane},\\Scikit-Learn~\cite{scikit-learn}\end{tabular} }                                                                      \\ \hline
\multicolumn{1}{|c|}{\cite{QFL-BlindC}}                          & \multicolumn{1}{c|}{\ding{56}}                                           & \multicolumn{1}{c|}{\begin{tabular}[c]{@{}c@{}} MNIST\cite{lecun1998gradient},\\ WDBC\cite{wdbc_uci} \end{tabular}}                                                                                                                              & \multicolumn{1}{c|}{5}                                           & \multicolumn{1}{c|}{200}                                                                                            & \multicolumn{1}{c|}{-}                                                                                              & \multicolumn{1}{c|}{\begin{tabular}[c]{@{}c@{}} 94\%\\ 98\% \\ \end{tabular}}                                                  & \multicolumn{1}{c|}{\begin{tabular}[c]{@{}c@{}} PennyLane~\cite{pennylane},\\ Qiskit~\cite{qiskit} \end{tabular}  }                                                                            \\ \hline
\multicolumn{1}{|c|}{\cite{QFL-DP}}                              & \multicolumn{1}{c|}{\ding{56}}                                           & \multicolumn{1}{c|}{Cats vs Dogs\cite{cats_vs_dogs_kaggle}}                                                                                                                             & \multicolumn{1}{c|}{100}                                         & \multicolumn{1}{c|}{-}                                                                                              & \multicolumn{1}{c|}{QFL-DP\cite{qfldp2021}}                                                                                         & \multicolumn{1}{c|}{98\%}                                                        & \multicolumn{1}{c|}{\begin{tabular}[c]{@{}c@{}} PyTorch~\cite{pytorch},\\ PyVacy~\cite{pyvacy} \end{tabular} }                                                                              \\ \hline
\multicolumn{1}{|c|}{\cite{QFL10}}                               & \multicolumn{1}{c|}{\ding{56}}                                           & \multicolumn{1}{c|}{\begin{tabular}[c]{@{}c@{}} MNIST\cite{lecun1998gradient},\\ CIFAR-10\cite{krizhevsky2009learning} \\ \end{tabular} }                                                                                                                           & \multicolumn{1}{c|}{3}                                           & \multicolumn{1}{c|}{50}                                                                                             & \multicolumn{1}{c|}{\begin{tabular}[c]{@{}c@{}}Quantum Secure \\ Aggregation\cite{QFL-SecAggPostQ1}\end{tabular}}                          & \multicolumn{1}{c|}{70\%}                                                        & \multicolumn{1}{c|}{-}                                                                                             \\ \hline
\multicolumn{1}{|c|}{\cite{gharavi2025pqbflpostquantumblockchainbasedprotocol}}                                                            & \multicolumn{1}{c|}{\ding{56}}                                              & \multicolumn{1}{c|}{-}                                                                                                                                         & \multicolumn{1}{c|}{-}                                            & \multicolumn{1}{c|}{150}                                                                                               & \multicolumn{1}{c|}{FedAvg \cite{mcmahan2017communication}}                                                                                               & \multicolumn{1}{c|}{-}                                                            & \multicolumn{1}{c|}{\begin{tabular}[c]{@{}c@{}}Post-Quantum\\Cryptography~\cite{postquantum}, \\ PyCryptodome~\cite{pycryptodome}\end{tabular}
}                                                                                             \\ \hline
\multicolumn{1}{|c|}{\cite{maouaki2025qfalquantumfederatedadversarial}}                                                            & \multicolumn{1}{c|}{\ding{56}}                                              & \multicolumn{1}{c|}{MNIST\cite{lecun1998gradient}}                                                                                                                                         & \multicolumn{1}{c|}{5, 10, 15}                                            & \multicolumn{1}{c|}{\begin{tabular}[c]{@{}c@{}}50(baseline), \\ 20(adversarial training)\end{tabular}}                                                                                               & \multicolumn{1}{c|}{FedAvg \cite{mcmahan2017communication}}                                                                                               & \multicolumn{1}{c|}{85\%}                                                            & \multicolumn{1}{c|}{-}                                                                                             \\ \hline
\multicolumn{1}{|c|}{\cite{tanbhir2025quantuminspiredprivacypreservingfederatedlearning}}                                                            & \multicolumn{1}{c|}{\ding{56}}                                              & \multicolumn{1}{c|}{OASIS MRI\cite{oasis_mri_marcus2007}}                                                                                                                                         & \multicolumn{1}{c|}{-}                                            & \multicolumn{1}{c|}{-}                                                                                               & \multicolumn{1}{c|}{FedAvg \cite{mcmahan2017communication}}                                                                                               & \multicolumn{1}{c|}{78\%}                                                            & \multicolumn{1}{c|}{-}                                                                                             \\ \hline
\multicolumn{1}{|c|}{\cite{10679217}}                                                            & \multicolumn{1}{c|}{\ding{56}}                                              & \multicolumn{1}{c|}{NSL-KDD\cite{NSL-KDD}}                                                                                                                                         & \multicolumn{1}{c|}{5-25}                                            & \multicolumn{1}{c|}{\begin{tabular}[c]{@{}c@{}} 25, \\ 50, 200 \end{tabular}}                                                                                               & \multicolumn{1}{c|}{FedAvg \cite{mcmahan2017communication}}                                                                                               & \multicolumn{1}{c|}{98\%}                                                            & \multicolumn{1}{c|}{PennyLane~\cite{pennylane}}                                                                                             \\ \hline
\multicolumn{1}{|c|}{\cite{DAQFL}}                                                            & \multicolumn{1}{c|}{\ding{52}}                                              & \multicolumn{1}{c|}{\begin{tabular}[c]{@{}c@{}}Fatal Health\cite{fetal_health},\\ Wisconsin Breast Cancer\cite{breast_cancer_wisconsin_diagnostic}\end{tabular}}                                                                                                                                         & \multicolumn{1}{c|}{2, 4, 8}                                            & \multicolumn{1}{c|}{-}                                                                                               & \multicolumn{1}{c|}{FedAvg \cite{mcmahan2017communication}}                                                                                               & \multicolumn{1}{c|}{93\%}                                                            & \multicolumn{1}{c|}{\begin{tabular}[c]{@{}c@{}}PyTorch~\cite{pytorch}, \\ PennyLane~\cite{pennylane}\end{tabular}}                                                                                             \\ \hline
\multicolumn{1}{|c|}{\cite{10829961}}                                                            & \multicolumn{1}{c|}{\ding{52}}                                              & \multicolumn{1}{c|}{\begin{tabular}[c]{@{}c@{}} MNIST\cite{lecun1998gradient}, \\ FMNIST\cite{fashion_mnist_xiao2017} \end{tabular}}                                                                                                                                         & \multicolumn{1}{c|}{-}                                            & \multicolumn{1}{c|}{100}                                                                                               & \multicolumn{1}{c|}{FedAvg \cite{mcmahan2017communication}}                                                                                               & \multicolumn{1}{c|}{52\%}                                                            & \multicolumn{1}{c|}{-}                                                                                             \\ \hline
\multicolumn{1}{|c|}{\cite{10878965}}                                                            & \multicolumn{1}{c|}{\ding{52}}                                              & \multicolumn{1}{c|}{mini-MNIST\cite{mini_mnist_no_official}}                                                                                                                                         & \multicolumn{1}{c|}{\begin{tabular}[c]{@{}c@{}}\\ - \\ \end{tabular}}                                            & \multicolumn{1}{c|}{100}                                                                                               & \multicolumn{1}{c|}{FedAvg \cite{mcmahan2017communication}}                                                                                               & \multicolumn{1}{c|}{52\%}                                                            & \multicolumn{1}{c|}{-}                \\ \hline
\multicolumn{1}{|c|}{\cite{10734228}}                                                            & \multicolumn{1}{c|}{\ding{56}}                                              & \multicolumn{1}{c|}{\begin{tabular}[c]{@{}c@{}}New England 39-bus\cite{new_england_39bus_athay1979} ,\\Nordic system VSA\cite{nordic_test_system_cigre}\end{tabular}}                                                                                                                                         & \multicolumn{1}{c|}{6}                                            & \multicolumn{1}{c|}{-}                                                                                               & \multicolumn{1}{c|}{FedAvg \cite{mcmahan2017communication}}                                                                                               & \multicolumn{1}{c|}{98\%}                                                            & \multicolumn{1}{c|}{-}                                          \\ \hline
\multicolumn{1}{|c|}{\cite{10636852}}                                                            & \multicolumn{1}{c|}{\ding{52}}                                              & \multicolumn{1}{c|}{\begin{tabular}[c]{@{}c@{}}CIFAR-10\cite{krizhevsky2009learning},\\MNIST\cite{lecun1998gradient}, FMNIST\cite{fashion_mnist_xiao2017},\\California Housing\cite{cal_housing_sklearn}\end{tabular}  }                                                                                                                                         & \multicolumn{1}{c|}{15}                                            & \multicolumn{1}{c|}{-}                                                                                               & \multicolumn{1}{c|}{FedAvg \cite{mcmahan2017communication}}                                                                                               & \multicolumn{1}{c|}{92\%}                                                            & \multicolumn{1}{c|}{\begin{tabular}[c]{@{}c@{}}D-Wave\\quantum annealer~\cite{dwave},\\Gurobi LP solver~\cite{gurobi}\end{tabular}}                                          \\ \hline

\multicolumn{1}{|c|}{\cite{qu2024qfsm}}                                                            & \multicolumn{1}{c|}{\ding{56}}                                              & \multicolumn{1}{c|}{\begin{tabular}[c]{@{}c@{}} CASIA\cite{casia_iris}, \\ RAVDESS\cite{livingstone2018ravdess}, EMO-DB\cite{emo_db_berlin} \end{tabular} }                                                                                                                                         & \multicolumn{1}{c|}{2, 4, 8}                                            & \multicolumn{1}{c|}{100}                                                                                               & \multicolumn{1}{c|}{FedAvg \cite{mcmahan2017communication}}                                                                                               & \multicolumn{1}{c|}{74\%}                                                            & \multicolumn{1}{c|}{PennyLane~\cite{pennylane}}                                                                             \\ \hline
\multicolumn{1}{|c|}{\cite{balasubramani2025novel}}                                                            & \multicolumn{1}{c|}{\ding{56}}                                              & \multicolumn{1}{c|}{NTUH\cite{ntuh_ecg_physionet}}                                                                                                                                         & \multicolumn{1}{c|}{-}                                            & \multicolumn{1}{c|}{-}                                                                                               & \multicolumn{1}{c|}{\begin{tabular}[c]{@{}c@{}}Secure\\Weighted Averaging \cite{Bonawitz2017SecAgg}\end{tabular}}                                                                                               & \multicolumn{1}{c|}{95\%}                                                            & \multicolumn{1}{c|}{\begin{tabular}[c]{@{}c@{}}Tensorflow Quantum~\cite{tfq},\\Tensorflow Federated~\cite{tff}\end{tabular}}   
\\ \hline
\multicolumn{1}{|c|}{\cite{10758814}}                                                            & \multicolumn{1}{c|}{\ding{52}}                                              & \multicolumn{1}{c|}{\begin{tabular}[c]{@{}c@{}}Didi Chuxing\\GAIA Initiative\cite{didi_gaia}\end{tabular}}                                                                                                                                         & \multicolumn{1}{c|}{100}                                             & \multicolumn{1}{c|}{50}  & \multicolumn{1}{c|}{\begin{tabular}[c]{@{}c@{}}Secure\\Weighted Averaging \cite{Bonawitz2017SecAgg}\end{tabular}}                                                                                               & \multicolumn{1}{c|}{95\%-98\%}                                                                                                                                                        & \multicolumn{1}{c|}{\begin{tabular}[c]{@{}c@{}}OpenAI Gym~\cite{openai_gym},\\Tensorflow Federated~\cite{tff},\\Tensorflow Quantum~\cite{tfq}\end{tabular}}                                                                               
\\ \hline

\end{tabular}
}
\label{tab:experimental_details}
\end{table*}

The experimental landscape in QFL exhibits substantial diversity in terms of data, number of clients, training rounds, aggregation methods, and performance outcomes.

\subsection{Experimental Details}
\begin{figure}[h]
    \centering
    \includegraphics[width=1\linewidth]{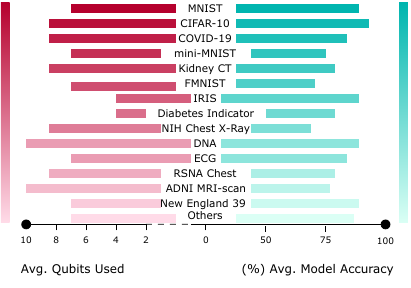}
    \caption{Reporting the datasets with their respective average number of qubits used and average model accuracy. The gradient (dark to light shades) signifies the total number of research available for each dataset.}
    \label{fig:datasets}
\end{figure}

Upon analyzing the experimental evaluations in the literature, we summarize some crucial details in Table~\ref{tab:experimental_details}. {\em MNIST}~\cite{lecun1998gradient} is observed as the most frequently used dataset, appearing in roughly one-third of studies; in contrast, specialized medical datasets (chest X-rays, MRI scans, etc.) represent an emerging application frontier. This preference for established benchmarks like MNIST suggests both practical convenience and a baseline foundation for comparability, even though it potentially limits exploration of quantum advantages in complex environments. {\em FedAvg}~\cite{mcmahan2017communication} dominates as the aggregation method, employed in approximately 70\% of implementations, reflecting its simplicity and effectiveness. However, it possibly constrains exploration of quantum-specific aggregation techniques. The {\em accuracy} metrics span impressively from modest 52\% (in non-IID mini-MNIST scenarios) to stellar 99\% achievements, with medical applications consistently surpassing 90\%. Yet these values resist meaningful comparison due to inconsistent experimental setups. Notably, only about 40\% of studies address {\em non-IID data} characteristics, leaving a critical gap given that data heterogeneity represents a fundamental challenge in real-world FL scenarios. Additionally, in Fig.~\ref{fig:datasets}, we report the utilization of qubits and model accuracy by different datasets commonly used in QFL frameworks. 

{\em Client participation} structures vary dramatically from minimal setups with 2--3 clients to expansive networks with 100. Most studies cluster around modest client counts from 5 to 15, suggesting computational limitations when scaling quantum circuits in federated environments. {\em Training rounds} show a similar variation (spanning from 15 to 100 rounds), indicating a practical balance between computational feasibility and model convergence. This pattern reveals a field that is still in its infancy between theoretical exploration and practical implementation constraints. The studies show increased attention to adversarial robustness (with dedicated training regimens requiring fewer rounds) and specialized applications like power systems and speech emotion recognition. Some studies achieve an impressive accuracy of 98\% with minimal client participation (3--5).

From an implementation aspect, the utilization of various {\em libraries} highlights the quantum ecosystem's maturation, with PennyLane~\cite{pennylane} emerging as the dominant framework (appearing in at least 30\% of specified implementations), followed by Tensorflow Quantum~\cite{tfq} and Qiskit~\cite{qiskit}. This technological evolution from theoretical proposals to practical implementations facilitates reproducibility.
The usage of D-Wave quantum annealer~\cite{dwave} in recent studies suggests growing interest in hardware-specific implementations beyond gate-based quantum computing. The striking accuracy disparity between conventional and quantum approaches on identical datasets raises important questions about quantum advantage in FL contexts - are these differences attributable to quantum properties, or do they merely reflect varying classical pre-processing techniques? 
Despite promising results, the lack of standardized benchmarking protocols remains a significant obstacle to meaningful cross-study comparisons and reliable assessment of quantum advantages in FL.

\begin{figure}[h]
    \centering
    \includegraphics[width=1\linewidth]{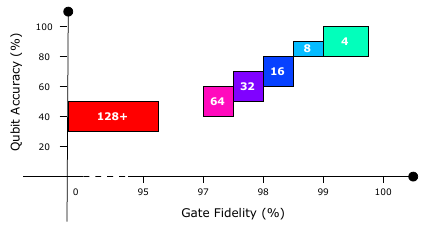}
    \caption{Gate fidelity versus accuracy with varying number of qubits. }
    \label{fig:QUE-analysis}
\end{figure}

\new{
Existing experimental studies on QFL studies are limited in scope, making statistical reliability crucial. Early frameworks validate feasibility with few clients and hybrid models\cite{QFL-TL-2}. Other works show that most QFL frameworks still operate with limited participants, shallow parametrized quantum circuits, synthetic partitions, and idealized conditions, leaving open questions about the necessary client count and conditions for statistical reliability in varied settings \cite{QFL-type, ballester_quantum_2025}. FL literature further indicates that client count, participation rate, and non-IID structure affect convergence and robustness highlights issues like label skew impacting simple aggregation techniques (such as FedAvg) \cite{FL1,hsu_non_iid_2019}. Recent reviews on data challenges in FL further emphasize that unreliable participation, skewed sample sizes, and unreported data distributions can cause overly optimistic results that fail to generalize and address heterogeneity \cite{saeed_datachallenges_2025,guendouzi_systematic_2023}. QFL inherits these issues and adds new challenges like shot noise, gate noise, device-specific error profiles, which together widen confidence intervals on performance metrics if the federation operates with too few or too homogeneous clients \cite{QFL-TL-2,ballester_quantum_2025}.
Current QFL experiments typically instantiate with several clients, often with synthetic non-IID splits and simplified quantum models limited by NISQ hardware limits. This choice is understandable from an implementation standpoint but it implies that many reported results are more feasibility demonstrations than proof of robustness at production scale. 

Few studies measure changes in accuracy, variance, and robustness to adversaries change as the number of participating nodes increases or as non-IID severity intensifies \cite{QFL-type,ballester_quantum_2025}. This is more challenging in QFL because quantum noise can mask or mimic heterogeneity effects, and small sample sizes on each device make  variance estimates less stable. A better approach would be to consider the minimum number of clients as a derived quantity that depends on heterogeneity and noise, not as an arbitrary hyperparameter. Reliability should focus on variance across clients, reporting local metric dispersion and how it changes with different sampling strategies. Classical FL shows increasing clients improves risk stability under non-IID settings \cite{FL1,hsu_non_iid_2019}. For QFL, this suggests that studies claiming strong generalization under heterogeneous conditions should (i) evaluate on configurations with numerous clients, (ii) use established non-IID protocols, and (iii) conduct multiple runs across random seeds to reveal outcome variations. 

Evaluations should report confidence bands, partitions, and device noise realizations to expose the spread of outcomes. Rather than focusing solely on point estimates ofnot just accuracy or loss, to highlight the effects of limited client counts and quantum noise. Non-IID data generally interacts with QFL-specific constraints in subtle ways. Strong label skew or feature bias amplifies the influence of a few high-capacity quantum clients if their updates dominate aggregation, particularly when these clients host deeper parametrized circuits or larger shot budgets. Hybrid QFL frameworks that adopt nested ansatz families or masked aggregation strategies mitigate this imbalance by sharing a common parameter core across all clients and restricting deeper blocks to capable devices, yet this design choice needs to be validated empirically with metrics that capture fairness and influence distribution across clients \cite{QFL-TL-2,FL1,ballester_quantum_2025}. Without such analysis, there is a risk that performance gains reported in small or symmetric testbeds conceal brittle behaviour under realistic deployment conditions where client capabilities and data distributions differ significantly.
}

Finally, along with ensuring practical environments for experiments, it is also to measure and analyze how well qubits have been utilized in a system, and what is their performance overall. In  Fig.~\ref{fig:QUE-analysis}, we illustrate the {\em Gate fidelity}~\cite{borgarino2025demystifying} along with qubit accuracy with varying number of qubit states. The gate fidelity quantifies how well the quantum gate performs what it is supposed to do, ideally without any errors. When the fidelity reaches 100\%, it implies the quantum gates are performing their intended operations without any errors, thus behaving ideally. For comparison, each box consists of the number of qubits utilized by the system. This comprehensive metric is vital since, despite high accuracy in individual qubits, low gate fidelity can accumulate errors throughout computations, thereby compromising the overall efficacy of quantum algorithms.

\subsection{Major Findings}

Our analysis of QFL frameworks uncovers major findings into the relationship between data heterogeneity, privacy techniques, and quantum-classical methods. 
\begin{itemize}
    \item Several studies highlight that non-IID data significantly hinders achieving high accuracy and stable convergence. For instance,~\cite{QFL-QAOA} shows a mere 81\% accuracy after 20 rounds in non-IID settings, underscoring the difficulty of integrating varied client updates efficiently. 
    \item The introduction of quantum elements into classical FL strategies has proven advantageous. Studies such as~\cite{QFL-TL-2} and~\cite{QFL-QNGD} provide compelling evidence that incorporating quantum circuits with classical optimization tactics (like FedAvg) enhances performance. Remarkably,~\cite{QFL-QNGD} achieves 99\% accuracy on MNIST using only 6 clients over 200 rounds, indicating that fewer clients suffice when iterative refinement is employed. 
    \item Privacy preservation techniques are pivotal in these analyses. Quantum security approaches, as illustrated in~\cite{QFL-DPFHE},~\cite{QFL-DPNoise}, and~\cite{QFL-BlindC}, succeed in maintaining high accuracy with minimal computational costs. For example,~\cite{QFL-DPFHE} mentions a 20\% reduction in overhead without compromising accuracy. These findings suggest that robust privacy implementations, when well-integrated, need not degrade efficiency.
    \item  Large-scale studies like those in \cite{QFL-QCNN} and \cite{QFL-TL-2} prove that QFL systems can scale from 10 to 100 clients while sustaining accuracy rates in the upper 90\% range, showcasing the system's ability to manage diverse client bases under challenging conditions. Emerging models, including the federated QLSTM in~\cite{QFL-QLSTM} and the quantum fuzzy neural network in \cite{QFL-QFuzzyNN}, expand QFL's reach beyond simple classification to include temporal data and fuzzy logic integration, potentially benefiting areas such as sensor networks and control systems. While these models may lack specific accuracy metrics, their swift convergence and adept handling of complex data suggest promising practical applications. 
\end{itemize}

\noindent Overall, the primary conclusions imply that achieving success with QFL frameworks involves balancing data diversity, iterative learning, and strong privacy strategies. These insights lay a solid foundation for further research to enhance QFL systems for varied, real-world environments.

\section{Discussion and Recommendations}
\label{sec:Discussion and Future Directions}

QC has the potential to revolutionize FL by addressing its core challenges and enabling novel applications across diverse domains. The inherent parallelism of quantum systems facilitates exponential speedups in solving complex optimization problems, such as those encountered in model training and aggregation, which are computationally intensive in classical FL systems. This advantage is particularly impactful in fields like drug discovery and material science, where QFL could enable faster and more accurate predictions on quantum-specific data. 
The ability of quantum devices to process high-dimensional data and operate effectively in noise-prone environments opens pathways for applications in decentralized sensing networks, autonomous systems, and large-scale industrial IoT platforms. As quantum hardware matures and hybrid quantum-classical systems become more practical, QFL frameworks can leverage the advancements to achieve greater scalability, robustness, and efficiency, driving innovation in AI and beyond.

\new{
While the addition of QC to QFL brings numerous advantages, it also generates qualitatively new concerns that researchers must consider while building novel QFL frameworks. QFL emphasizes enhanced model inversion via quantum expressivity, allowing parametrized quantum circuits to offer richer hypotheses than classical models in certain tasks. This expressivity, combined with structured measurement statistics, can enhance the link between updates and client data. An adversary equipped with quantum or high-performance classical solvers may run reconstruction attacks, motivating attack risks from quantum-capable adversaries, warranting robust security definitions \cite{zhao_federation_2025,carletti_sok_2025}. QFL also introduces threats like quantum channel and entanglement abuse, where compromised coordinators might exploit quantum networks to manipulate client systems beyond intended protocols. Existing QFL taxonomies do not capture this nonlocal manipulation capability. Recent QFL surveys already emphasize that security analyses rarely model malicious entanglement or untrusted quantum switches, indicating an open gap between protocol design and adversarial modeling \cite{ballester_quantum_2025,nguyen_qfl_survey_2025}. 

Further, delegated QC in QFL merges cloud threats with FL assumptions, risking data integrity and confidentiality. A malicious quantum service can bias gradients, embed structured errors that evade simple norm-based anomaly detection, or harvest side-channel information about client circuits and data encodings. This behaviour blurs the boundary between model poisoning and hardware-level attacks and require joint analysis of delegated computation security and FL robust aggregation. The drive towards post-quantum cryptography and quantum-safe secure aggregation itself becomes a potential attack surface. Recent works on post quantum-secure FL aggregation demonstrate that careful parameterization is needed to avoid new leakage patterns or degradation in robustness~\cite{QFL-SecAggPostQ2,Zhang2025_SAReview}. In QFL, where all adversaries are, by construction, quantum-capable, flawed or partial deployments of post-quantum primitives can create a false sense of protection while leaving update streams exploitable.
}

On the threats aspect, while a broad spectrum of threats is conceivable in FL, early deployments of QFL necessitate a careful prioritization of those most likely to impair real-world pilots. The distinctive constraints of the NISQ era imply that certain attack vectors pose disproportionately higher risks~\cite{baseri2025futureproofingcloudsecurityquantum}.  
First, poisoning attacks on model updates and data inputs are a major concern. As in FL, even a few malicious updates can degrade the global model. This risk is greater in QFL due to limited qubits and high measurement variance, complicating the differentiation between benign noise and adversarial perturbations. Second, protecting client information through gradient updates or quantum states is crucial. Hybrid quantum-classical processing might expose sensitive data through transmitted gradients or reduced states. 
Third, protecting against untrusted computation nodes like cloud QPU providers or aggregation servers is critical. Honest-but-curious participants may try to deduce client contributions, emphasizing the need for quantum-secure multi-party protocols with minimal overhead. Other classes of threats, such as collusion between semi-honest nodes or gate parameter manipulation in large-scale quantum devices, are of medium to lower priority in the near term. While important for the long-term roadmap, their impact is constrained by current hardware capabilities and deployment scales. Addressing all of these with minimal overhead under NISQ constraints can form the foundation for secure and trustworthy QFL systems.

\begin{figure}[h]
    \centering
    \includegraphics[width=0.88\linewidth]{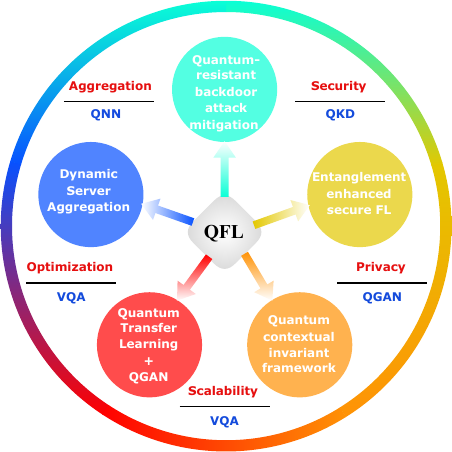}
    \caption{Highlighting key research areas along with identified gaps (in red-colored text) for future work with their possible solution direction (in blue-colored text). 
    }
    \label{fig:key_area}
\end{figure}

The field of QFL continues to expand, with numerous challenges and opportunities driving research in this domain. This section critically explores unresolved issues and highlights promising directions for future work, structured across key areas of exploration, as illustrated in Fig.~\ref{fig:key_area}. 

\noindent $\bullet$ \textit{\textbf{Dynamic server aggregation}}: 
Centralized aggregation exposes QFL to trust concentration and a single point of failure. A future direction could be a rotating server committee, selected in each round by randomness that is both unpredictable and publicly verifiable~\cite{HerreroCollantes2017QRNG,Liu2018DIQRNG}. Quantum random number generators and their device-independent variants can provide certified entropy, while key derivation functions transform that entropy into selection seeds. Adding these seeds generate auditable vectors that specify which servers and clients participate in aggregation, and their publication through cryptographic techniques allows independent verification of the process. Further, threshold secure aggregation must then operate across the rotating committee, with explicit parameters for quorum, dropout recovery, and rekeying during rotation~\cite{Bonawitz2017SecAgg,Bell2020SecAggPlus}. Designing such a framework distributes authority, curbs collusion, and introduces accountability, while also demanding systematic evaluation of its latency, privacy retention, and fairness relative to fixed-server baselines. 

\noindent $\bullet$ \textit{\textbf{Quantum transfer learning with quantum generative adversarial networks}}: 
Knowledge transfer across quantum models is limited by architectural differences in encoders, ans\"atze, and readout, which leads to incompatible feature representations across devices. A practical way forward is to use quantum generative adversarial networks (QGANs)~\cite{dallaire2018quantum,ngo2023survey} to learn transferable features that bridge these mismatches. Concretely, one can train generator–discriminator pairs to translate measured feature distributions from a source architecture to a target one, so downstream tasks on the target benefit from the structure learned on the source. This translator can be coupled with a lightweight quantum variational autoencoder module to regularize the latent space and stabilize training, while the GAN objective focuses on distribution alignment. The resulting framework yields a flexible transfer pipeline: learn source features, map them through the QGAN translator to the target representation, then fine-tune on the target with minimal new data. Such a setup can support cross-device reuse in QFL where client hardware is heterogeneous, and opens a clear research track on stable, device-agnostic feature translation with rigorous evaluation under NISQ constraints.

\noindent $\bullet$ \textit{\textbf{Quantum contextual invariant robustness framework}}: 
A persistent challenge in distributed learning is the degradation of stability when models operate across heterogeneous environments. Future research can address this by developing quantum circuits that exploit symmetry-preserving operations to maintain invariants during training and communication. Contextual transformations, when encoded as unitary symmetries, can be preserved through quantum contextuality, providing robustness against distortions induced by noise or cross-platform variability~\cite{Howard2014Contextuality,Raussendorf2013Contextuality}. Integrating these invariants with federated aggregation may allow models to retain consistency without retraining overhead, offering a structured pathway to improve cross-device transfer in QFL. The open question lies in quantifying how such invariants can be efficiently implemented on NISQ devices, where gate noise and limited qubits impose strict constraints on symmetry exploitation.

\noindent $\bullet$ \textit{\textbf{Quantum entanglement-enhanced secure FL}}:
Privacy in FL under classical secure aggregation is vulnerable due to collusion or side-channel leaks. A stronger method is to use the entanglement-assisted QKD, providing information-theoretic secure keys for encryption with one-time pads in every round, reducing exposure even with compromised relays~\cite{Ekert1991E91,Lo2012MDIQKD}. Entanglement allows measurement-device–independent protocols, eliminating reliance on trusted detectors, addressing loopholes in standard QKD~\cite{Scarani2009QKDRev,Pirandola2020QKDReview}. This maintains the classical nature of model updates while enhancing cryptographic strength~\cite{Wootters1982NoCloning}. While limitations include finite key rates, distance constraints, and reliance on underdeveloped quantum repeaters~\cite{Wehner2018QI}, future work should assess partial-participation protocols under finite-key conditions and compare against forward-secure post-quantum cryptography to determine the benefits of entanglement-based integration.

\noindent $\bullet$ \textit{\textbf{Quantum-resistant backdoor attack mitigation framework}}:
Backdoor attacks remain a core risk in FL and extend to QFL through model replacement and distributed trigger injection~\cite{bagdasaryan2020how,xie2020dba}. Classical defenses rely on static anomaly detection, which often fails under distributed and adaptive poisoning strategies. A more promising direction is to combine robust aggregation with quantum subroutines that expose subtle correlations hidden from classical inspection. Swap-test circuits can compare suspect updates against trusted references to reveal anomalous overlaps~\cite{paris2004qse}, while classical-shadow tomography enables efficient screening of many observables from few measurements~\cite{huang2020shadows}. These tools shift detection from coarse statistical checks to fine-grained quantum probes, providing resilience even in low-data regimes. The challenge remains to integrate them without excessive overhead on NISQ devices, motivating hybrid pipelines that use lightweight classical filters and escalate only confirmed cases to quantum-level scrutiny.

\section*{Acknowledgment}
The work is supported by BITS Pilani Dubai Campus (BPDC) for the project ``FaR-FedIoT: Fair and Resource-adaptive Federated System
for IoT" under NFSG scheme.

\bibliography{references}
\newpage
\vspace{-0.3cm}
\begin{IEEEbiography}[{\includegraphics[width=1in,height=1.25in, clip]{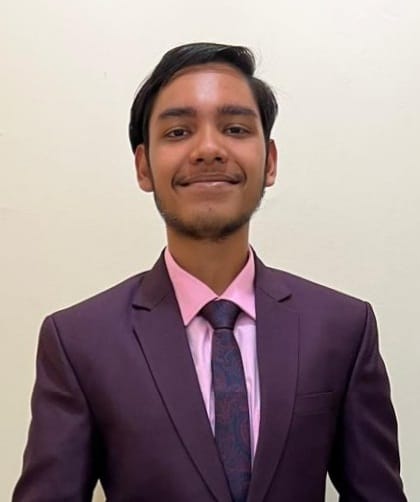}}]{Aakar Mathur} is a senior undergraduate in Computer Science at BITS Pilani, Dubai Campus, UAE. He has worked on hands-on machine learning models at Emirates Airlines and has gained practical data analysis experience at Apparel Group. He has been working as a web development intern since 2025 at Siemens Energy, Dubai. His research interests include quantum computing, federated learning, machine \& deep learning, and behavioural finance. 
\end{IEEEbiography}

\vspace{-2cm}
\begin{IEEEbiography}[{\includegraphics[width=1in,height=1.25in, clip]{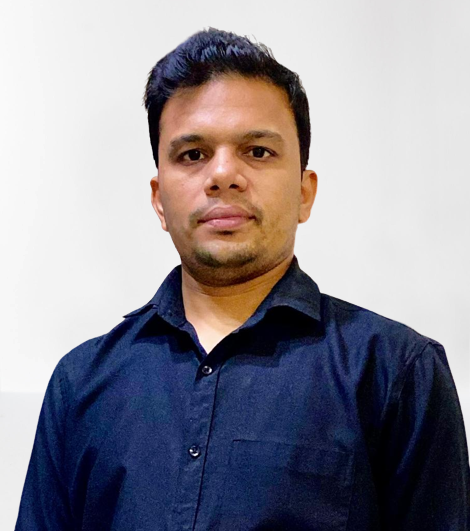}}]{Ashish Gupta} is an Assistant Professor in Computer Science at BITS Pilani Dubai Campus, UAE. He has been working as a postdoctoral fellow in the Department of Computer Science at Missouri S\&T, USA, from 2021 to 2023. He received a Ph.D. in Computer Science and Engineering from the Indian Institute of Technology (BHU), Varanasi, India. His research interests include sensor data analytics, federated learning, and applied machine learning. 
\end{IEEEbiography}

\vspace{-2cm}
\begin{IEEEbiography}[{\includegraphics[width=1in,height=1.25in, clip]{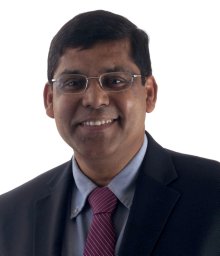}}]{Sajal K. Das} (Fellow, IEEE) is a Curator's Distinguished Professor of Computer Science and the Daniel St. Clair Endowed Chair with Missouri University of Science and Technology, USA. His research interests include wireless and sensor networks, mobile and pervasive computing, mobile crowdsensing, cyber-physical systems and IoT, cloud and edge computing, cyber security, and social networks. He has co-authored more than 700 research articles in high-quality journals and refereed conference proceedings, five U.S. patents, and four books. His h-index is 100+ with more than 45,000 citations. 
He has been the Editor-in-Chief of Pervasive and Mobile Computing Journal (Elsevier’s) and an Associate Editor of IEEE Transactions on Dependable and Secure Computing. 
\end{IEEEbiography}
\end{document}